\documentclass[preprint,12pt]{elsarticle}
\usepackage{amssymb}
\usepackage[utf8]{inputenc}
\usepackage[english]{babel}
\usepackage{alphabeta} 
\usepackage{amsmath}
\usepackage{fullpage}
\usepackage{anyfontsize}
\usepackage{setspace}
\usepackage{color}
\usepackage{scrextend}
\usepackage{boldline}
\usepackage{graphicx}
\usepackage{changepage}
\graphicspath{ {Images/} }
\usepackage{pagecolor,lipsum}
\usepackage{url}
\usepackage{siunitx}
\usepackage{multicol}
\usepackage{multirow}
\usepackage[top=1in, bottom=1in, left=0.94in, right=0.94in]{geometry}
\usepackage[acronym]{glossaries}
\usepackage[toc,page]{appendix}
\newcommand{\eat}[1]{}

\usepackage{subcaption}
\usepackage[official]{eurosym}
\usepackage{enumitem}
\setlist{nolistsep,noitemsep}
\usepackage[hidelinks]{hyperref}
\usepackage{cleveref}
\crefname{figure}{Figure}{Figures}
\crefname{equation}{Equation}{Equations}
\usepackage{caption}
\usepackage{lipsum}
\usepackage{bm}
\DeclareUnicodeCharacter{2212}{-}
\newcounter{daggerfootnote}

\journal{Astroparticle Physics}

\begin{document}
\begin{frontmatter}
\title{Gamma-ray background from rock: studies for a next-generation
dark matter experiment based on liquid xenon}
\author[inst1,inst2]{J. K. Tranter}

\affiliation[inst1]{organization={Department of Physics and Astronomy, The University of Sheffield}, 
            city={Sheffield},
            postcode={S3 7RH},
            country={UK}}

\author[inst1]{V. A. Kudryavtsev}
\author[inst2]{P. R. Scovell}

\affiliation[inst2]{organization={STFC, Boulby Underground Laboratory},
            addressline={Boulby Mine},
            city={Redcar-and-Cleveland},
            postcode={TS13 4UZ},
            country={UK}}

\begin{abstract}
Rare event experiments, such as those targeting dark matter interactions and neutrinoless double beta (0νββ) decay, should be shielded from gamma-rays that originated in rock. This paper describes the simulation of gamma-ray transport through the water shielding and assessment of the thickness needed to suppress the background from rock down to a negligible level. This study focuses on a next-generation xenon observatory with a wide range of measurements including the search for Weakly Interacting Massive Particles (WIMPs) and 0νββ decay of $^{136}$Xe. Our findings indicate that the gamma-ray background is unlikely to persist through analysis cuts in the WIMP energy range (0~-~20~keV) after 3.5~m of water, complemented by 0.5~m of liquid scintillator. For 0νββ decay, a background below 1 event in 10 years of running can be achieved with a fiducial mass of 39.3~tonnes. Furthermore, for typical radioactivity levels of 1~Bq~kg$^{-1}$ of $^{232}$Th and $^{238}$U we have studied the effect of reducing the water shielding by 1~m, resulting in a reduced fiducial mass of 19.1~tonnes for 0νββ decay and still a negligible background for WIMP search. The paper also presents the measurements of radioactivity in rock in the Boulby mine, which hosted several dark matter experiments in the past and is also a potential site for a future dual-phase xenon experiment. The measurements are used to normalise simulation results and assess the required shielding at Boulby.
\end{abstract}

\end{frontmatter}

\section{Introduction}
\label{sec:intro}
We consider the next-generation dark matter detector based on dual-phase xenon technology, with the primary aim of probing for Weakly Interacting Massive Particle (WIMP) interactions. Recent results from current liquid xenon (LXe) detectors searching for WIMPs with masses ranging from a few GeV/c$^2$ to tens of TeV/c$^2$ have established the most stringent limits on spin-independent cross-sections~\cite{LZfirstResults,PandaX,XENONnT}. The sensitivity of the future detector is projected to be more than a magnitude better than current limits, down to a spin-independent cross-section of $3\times10^{-49}$~cm$^2$ at 40~GeV/c$^2$~\cite{XLZD}. The detector will also be able to conduct a range of other low-background searches. Since LXe contains the isotope $^{136}$Xe, it will be possible to search for evidence of neutrinoless double beta (0νββ) decay, complementing other experiments dedicated to finding the Majorana nature of neutrinos, with LXe and germanium~\cite{0vbb_review}. 

One of the critical challenges for the success of the next-generation experiment is minimising the background. Background events can be caused by various processes involving radioactivity and cosmic rays. Decay chains of $^{238, 235}$U and $^{232}$Th together with $^{40}$K are prime sources of radioactivity. Alphas, betas, gamma-rays and neutrons (from spontaneous fission and $(\alpha,n)$ reactions) produced by these nuclides and their progeny can cause events that mimic WIMP interactions and $0\nu\beta\beta$ decay if their energy depositions are reconstructed as single scatters within the detector fiducial volume and the energy region of interest (ROI), and cannot be discriminated from the signal by any analysis technique. These requirements leave only gamma-rays and neutrons as background-inducing particles if the radioactive nuclides are located sufficiently far from the target. Alphas and betas, as background-causing events, are limited to the target and nearby components. Gaseous, radioactive radon emanating from detector components in contact with the target material (LXe in our case) may induce background events via $^{214}$Pb decays to the ground state producing betas with an energy spectrum extending down to the WIMP search ROI. Screening materials for radioactivity, selecting those with minimum radioactive contaminants, and cleaning surfaces of components in contact with the target help suppress background radiations. Some materials may also contain non-negligible contaminations from $^{60}$Co and other activated nuclides. The rate of cosmic-ray muons is suppressed by several orders of magnitude in deep underground locations. Additional suppression of the muon-induced background is provided by veto systems that allow signal rejection in (delayed) time coincidence with a muon.

A range of monoenergetic gamma-rays are emitted during the radioactive decays of the nuclides in uranium and thorium decay chains in rock. As they travel through water shielding and detector materials, most gamma-rays will lose some or all of their energy via various processes, such as Compton scattering and photoelectric absorption. By the time they reach the detector, there is a large spectrum of gamma-rays with energies from 0~-~2.615~MeV. The upper limit here is set by the highest energy gamma-ray with high intensity: $^{208}$Tl decaying to $^{208}$Pb with an intensity of 99.754\,\%, late in the $^{232}$Th decay chain~\cite{Nudat_Tl208}.

The shielding of the detector must be thick enough to deter the majority of these gamma-rays such that after stringent analysis cuts, there is $<$~1 event per year in the WIMP energy ROI, and $<$~1 event per 10~years around the 0νββ decay Q-value. For WIMPs, a powerful discrimination of $>$~99.5\,\% between nuclear recoils (NRs) from WIMPs and electron recoils (ERs) from electrons and gamma-rays will suppress the background from rock gamma-rays further by at least two orders of magnitude~\cite{LZfirstResults}. For 0νββ decay, there is no further suppression as signal events will look almost exactly like gamma-induced electrons.
It should be noted that neutrons from spontaneous fission and $(\alpha, n)$ reactions in the rock are attenuated quicker in water than gamma-rays. Therefore, we assume that shielding thickness sufficient to reduce the gamma-ray background will be adequate for rock neutrons as well. Section~\ref{sec:simulation} describes the establishment and results of a simulation developed to investigate the shielding thickness.

There is a potential for the future detector to be housed in Boulby Mine, North Yorkshire, UK. The site where the future cavern would be built is approximately 1300~m below sea level~\cite{1300m}, 200~m deeper than where the current Boulby Underground Laboratory is situated~\cite{BUL}. Rock samples from both 1100~m and 1300~m level sites in the mine have been retrieved by ICL mining company via boreholes. They were given to the Boulby Underground Laboratory to be screened by sensitive, high-purity germanium (HPGe) detectors in the Boulby UnderGround Screening (BUGS) facility~\cite{BUGS}. The results of this rock sample screening for gamma-rays are presented in Section~\ref{sec:boulby}, intended for use by any prospective experiments that Boulby Mine may host. Further considerations regarding the effects of a reduction in water shielding on the background to WIMP and 0νββ decay searches are discussed in Section~\ref{sec:discussion}.

\section{GEANT4 simulation of gamma-rays from rock}
\label{sec:simulation}
To investigate the gamma-ray background from rock in the detector and determine sufficient shielding thickness, a simulation has been developed in GEANT4~\cite{GEANT4}. The code generates gamma-rays in a layer of rock surrounding a large, cylindrical cavern and propagates them towards the detector surrounded by a water tank and built on top of a steel plate.
As there is currently no designated location for a future detector, results from the simulation have been normalised here to 1~Bq\,kg$^{-1}$ of radiation for $^{238}$U, $^{232}$Th and $^{40}$K. Therefore, the background event rate can be scaled with the measured activity of these radionuclides for a specific cavern and detector location. Similar work by the LZ collaboration has been done to look at the contribution of gamma-rays from surrounding rock to the ER background~\cite{LZ_sim}. Simulations such as these inform experimentalists of the required thickness of shielding so the cavern can be designed to accommodate all systems, including shielding. The simulation in this report was created with GEANT4 version 10.7.2 and the relevant standard physics list ``Shielding'' was used.

\subsection{Simulation geometry}
\label{sec:sim_geom}
\begin{figure}[htbp] 
    \begin{subfigure}{.46\textwidth}
    \centering
    \centering\includegraphics[width=1\textwidth,angle=0]{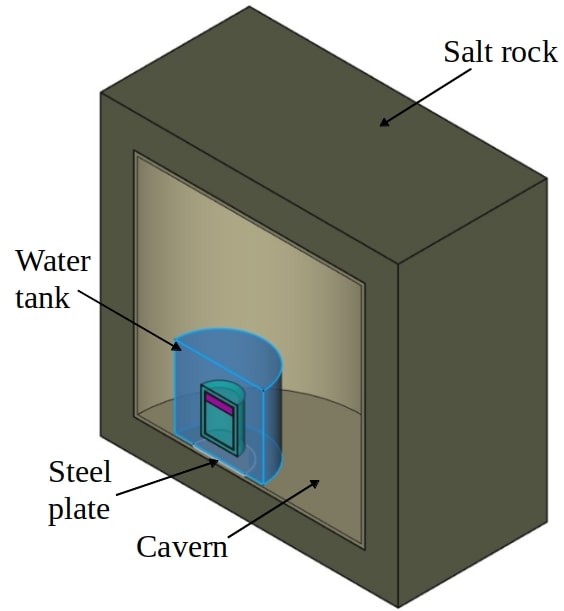}
        \caption{}
    \label{fig:geom_a}
    \end{subfigure}
    \begin{subfigure}{.48\textwidth}
    \centering
    \centering\includegraphics[width=1\textwidth,angle=0]{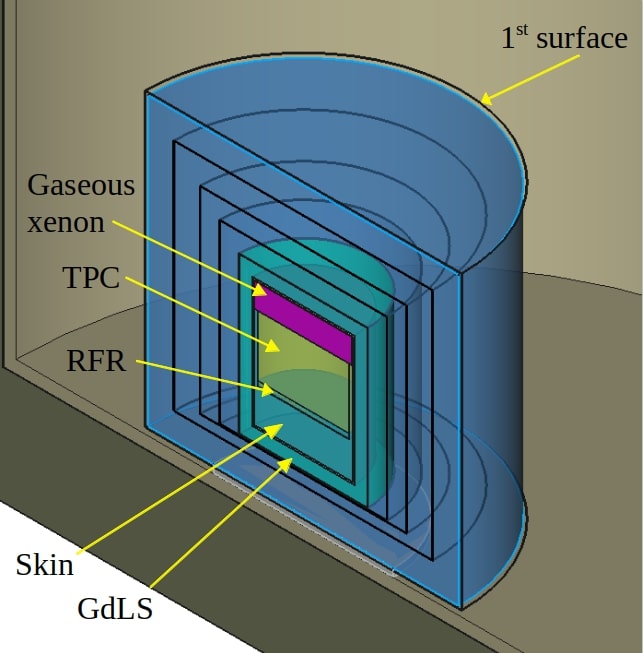}
        \caption{}
    \label{fig:geom_b}
    \end{subfigure}
    \caption{\small{(a) A 3D cross-sectional schematic diagram of the simulated rock, cavern, water tank and detector. (b) A closer look at the internal structure of the LXe detector, surrounded by the water tank and the 50~cm thick rock layer from which the gamma-rays were generated. The thick, black lines represent the surfaces that the gamma-rays were stopped at and re-propagated from.}}
    \label{fig:geom}
\end{figure}

The simulation's geometry, similar to that described in Ref.~\cite{Vpec}, has been developed based on the scaled-up design of the LZ detector~\cite{LZ}. The model is shown in \autoref{fig:geom} and includes a cylindrical cavern of 30~m height and 30~m diameter surrounded by salt rock. The detector sits at the bottom of the cavern, offset from the centre, and consists of a titanium, double-walled vacuum cryostat surrounded by 0.5~m of Gd-loaded liquid scintillator (GdLS) and a water tank. The walls of the cryostat are 2~cm thick with 5~cm of vacuum separating them. No material separates the GdLS from the water or the water from the cavern air in this simulation. The cryostat houses a dual-phase time projection chamber (TPC) lined with PTFE and filled with LXe, providing an active mass of 71~t with a density of 2.953~g~cm$^{-3}$. There is a small volume of gaseous xenon (GXe) on top of the LXe and an additional layer of LXe below the active volume which mimics the reverse field region (RFR) in a real dual-phase xenon detector. There is also a layer of LXe around the TPC and below it, called the ``skin'', which acts as a gamma-ray veto. In the actual construction of the detector, the gate and anode grids will be at the base and top of the GXe dome, respectively, while the cathode grid will be above the RFR. However, no structures or materials for any of the grids are included in the simulation. Changes in geometry that could affect the results of this study would be thickness variation of high-density materials such as LXe and titanium. For instance, increasing (decreasing) the thickness of titanium by 50~\% would decrease (increase) the event rate in the TPC by a factor of $\sim1.4$.

The GdLS is used as a veto primarily for neutron interactions in the TPC due to a high thermal neutron capture cross-section on $^{157}$Gd. Both the skin and GdLS allow the rejection of gamma- and neutron-induced events that may mimic a WIMP signal in the TPC. There is 3.5~m of water shielding on the sides and top of the GdLS layer, and 1.5~m of water below the GdLS layer. The overall height of the water tank is 10.94~m with a diameter of 11.90~m. There is a 30~cm thick stainless steel plate underneath the water tank which allows us to reduce the required thickness of water and hence, the height of the support structure for the TPC and GdLS tanks, but can be replaced by a similar thickness of water as on the top and sides.

To help with the task of simulating gamma-rays from the cavern rock, extra volumes were added around the detector and cavern: a 50~cm thick rock layer, within which gamma-rays were generated, was placed around the cavern volume, and ``parallel world'' surfaces (hereafter named 1st surface, 2nd surface and so on) were placed around and concentrically throughout the water tank. These surfaces are a useful tracking feature of GEANT4 as they can overlap with ``real'' volumes without intervening in particle transport, however, we can still extract information about the particles as they cross the surfaces. This feature allows the number of simulated gamma-rays to be boosted in a multi-stage process, further outlined in Section~\ref{sec:gamma_prod}, which is required because over several trillion gamma-rays need to be generated to obtain statistically acceptable data in the detector after they are attenuated through shielding, and this is computationally unattainable. The surfaces are separated as listed in \autoref{tab:surfaces}. \autoref{fig:geom_b} displays the geometric sections in detail, including these added surfaces.

\begin{table}[htbp]
\def\arraystretch{1.1}
\caption{The geometrical parameters of the tracking surfaces within the GEANT4 simulation. The water tank (WT), GdLS, and outer cryostat vessel (OCV) volumes are included for reference. The last two columns give distances subtracted inwards from the edges of the water tank.}
\vspace{-5 mm}
\label{tab:surfaces}
\begin{center}
\begin{tabular}{|c|c|c|c|c|}
\hline
\multicolumn{1}{|c|}{\textbf{Volume}} & \multicolumn{1}{c|}{\textbf{Height [m]}} & \multicolumn{1}{c|}{\textbf{Diameter [m]}} & \multicolumn{2}{c|}{\textbf{Distance from WT [m]}} \\
\cline{4-5}
\multicolumn{1}{|c|}{\textbf{}} & \multicolumn{1}{c|}{\textbf{}} & \multicolumn{1}{c|}{\textbf{}} & \multicolumn{1}{c|}{\textbf{Top \& sides}} & \multicolumn{1}{c|}{\textbf{Base}} \\
 \hline
1st surface& 11.14 & 12.10 & -0.10 & -0.10 \\
WT & 10.94 & 11.90 & 0 & 0 \\
2nd surface& 8.94 & 9.90 & 1.00 & 1.00 \\
3rd surface& 7.54 & 7.90 & 2.00 & 1.40 \\
4th surface& 6.74 & 6.40 & 2.75 & 1.45 \\
5th surface& 5.96 & 4.92 & 3.49 & 1.49 \\
GdLS & 5.94 & 4.90 & 3.50 & 1.50 \\
6th surface & 4.96 & 3.92 & 3.99 & 1.99 \\
OCV & 4.94 & 3.90 & 4.00 & 2.00 \\
\hline
\end{tabular}
\end{center}
\end{table}

\subsection{Gamma-ray production}
\label{sec:gamma_prod}
\begin{figure}[hbtp] 
    \begin{subfigure}{.49\textwidth}
    \centering
    \centering\includegraphics[width=1\textwidth,angle=0]{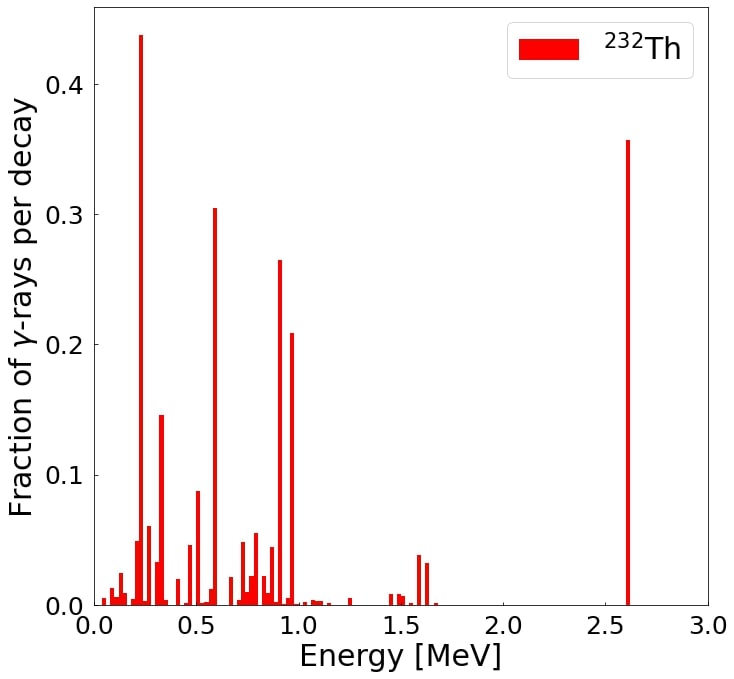}
        \caption{}
    \label{fig:Initial_energyTh}
    \end{subfigure}
    \begin{subfigure}{.49\textwidth}
    \centering
    \centering\includegraphics[width=1\textwidth,angle=0]{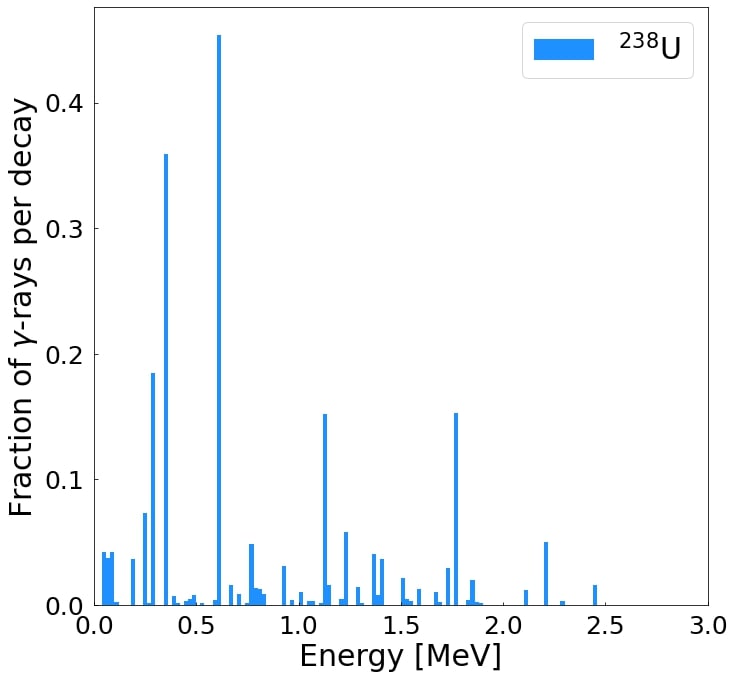}
        \caption{}
    \label{fig:Initial_energyU}
    \end{subfigure}
    \caption{Initial energy spectra of $10^{6}$~gamma-rays for (a) the $^{232}$Th decay chain and (b) the $^{238}$U decay chain, normalised to each parent nuclide decay.}
    \label{fig:init_energies}
\end{figure}

Gamma-rays were simulated as individual photons with energies and intensities. Coincident gamma-rays were not considered as they can be neglected when the source is far away from the target. To replicate $^{232}$Th and $^{238}$U activity, gamma lines with high intensities from their respective decay chains were generated in the rock volume around the cavern. Accounting for only these gamma lines, the average number of gamma-rays emitted per parent decay, $N_{\gamma}$, is 2.4567 for $^{232}$Th and 2.0955 for $^{238}$U. The intensity of each of these gamma lines is expressed as a fraction of $N_{\gamma}$ for the decay chain they belong to. $10^6$~gamma-rays have been generated with GEANT4 and their intensities have been plotted in \autoref{fig:init_energies}. The fraction of gamma-rays in each line agrees very well with gamma-ray intensities from NNDC~\cite{Nudat_Tl208}, proving that there is no bias in the simulation at the particle generation stage. It was initially assumed that the activity of the parent radionuclides was 1~Bq~kg$^{-1}$.

\par The isotope $^{40}$K decays to $^{40}$Ar with a branching ratio of 10.72\,\% via electron capture, with the emission of a neutrino and a 1.461~MeV gamma-ray~\cite{Nudat_K40}, hence, we generated this decay as a monoenergetic gamma-ray. In one of the simulations, we also generated a monoenergetic gamma line of 2.615~MeV from $^{208}$Tl decay (99.7\,\% intensity)~\cite{Nudat_Tl208}. $^{208}$Tl is the result of $^{212}$Bi alpha decay with a 35.94\,\% branching ratio. The purpose of this particular simulation was to compare the transport of the gamma-rays with the highest energy (2.615~MeV) with the whole $^{232}$Th decay chain spectrum. This comparison is discussed further in Section~\ref{sec:gamma_prop}.

The rock layer's thickness of 50~cm to generate gamma-rays was chosen as it was estimated that gamma-rays which started more than 50~cm into the rock would be heavily attenuated, thereby providing insignificant statistics compared to gamma-rays closer to the cavern. To validate this assumption, different thicknesses were tested to see the change in the percentage of gamma-rays reaching the first surface. When generated within a 50~cm thick rock layer, 98.3\,\% of the gamma-rays that reached the surface originated from the first 40~cm into the rock. Hence, a 50~cm layer was considered further on as producing almost all gamma-rays reaching the cavern.

\subsection{Gamma-ray propagation through the shielding and energy deposition in the TPC}
\label{sec:gamma_prop}
Once generated, the gamma-rays are transported through the cavern and shielding towards the TPC. They are stopped at each of the concentric surfaces before being propagated again towards the next surface in increased numbers. This is a biasing method, previously used in the LZ experiment's simulation campaign and documented in Ref~\cite{LZ_sim}. The initial number of gamma-rays is increased by a factor $F$:
\begin{align}\label{eq:eq1}
    F = \prod_{i=1}^{n} m_i
\end{align}
where $m_i$ represents the multiplication factor by which each gamma-ray is re-propagated and $n$ is the number of surfaces. This allows the user to boost statistics by enlarging the total number of gamma-rays generated. $F$ is a statistical factor for all events; no additional statistical weights have been applied. The energy, position and momentum direction at each surface are stored. The values of $F$ for $^{232}$Th, $^{238}$U, and $^{40}$K are, respectively: $1.8\times10^{9}$, $5.1\times10^{9}$ and $2.5\times10^{9}$. 

\begin{figure}[ht!]
    \centering
    \centering\includegraphics[width=0.8\textwidth,angle=0]{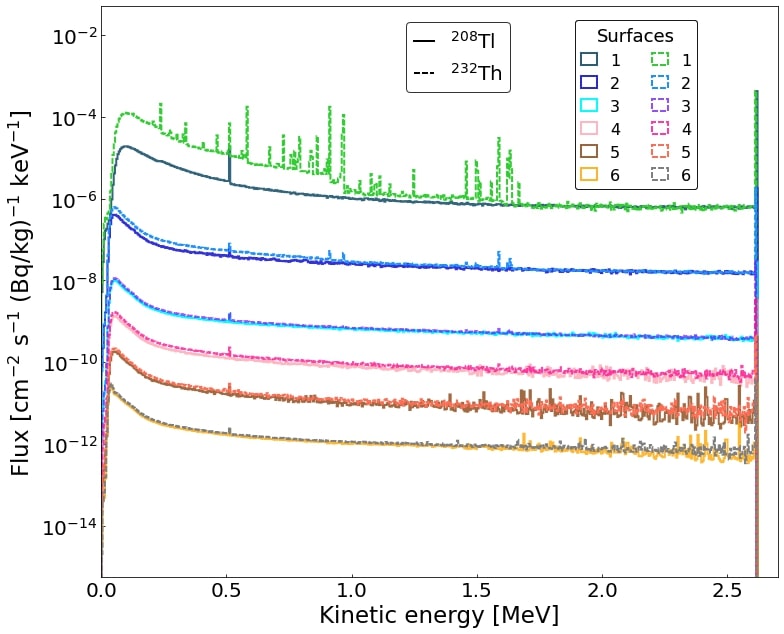}
        \caption{Gamma-ray energy spectra from $^{208}$Tl (solid line) and the whole chain of $^{232}$Th (dashed line) decays at each surface from the outside of the water tank (surface 1) and in stages throughout until reaching the outside of the TPC (surface 6). The surfaces through the shielding are illustrated in \autoref{fig:geom_b} and the distances between each of these surfaces and certain detector volumes are listed in \autoref{tab:surfaces}. The peak at 0.511~MeV is due to electron-positron annihilation, where positrons are produced via the pair production process by high-energy gamma-rays in the water.}
    \label{fig:attenuation}
\end{figure}

The decay rate of the parent radionuclide was normalised to 1~Bq~kg$^{-1}$, and the energy spectrum of gamma-rays at each surface was expressed in counts cm$^{-2}$~s$^{-1}$~(Bq/kg)$^{-1}$~keV$^{-1}$. \autoref{fig:attenuation} shows the energy spectra of gamma-rays at each surface from decays of the nuclides $^{232}$Th (the whole decay chain) and $^{208}$Tl. It shows that by the time the gamma-rays reach the TPC, only those originating from $^{208}$Tl remain, meaning that the $^{208}$Tl and $^{232}$Th spectra are very similar. This occurs due to the higher attenuation coefficient, $A$, for lower-energy gamma-rays. For completeness, the full $^{232}$Th spectrum was used for the final analysis in Section~\ref{sec:analysis}.

According to a separate, simple GEANT4 simulation, the attenuation coefficient of a 2.615~MeV gamma-ray through water is 0.0426~cm$^{2}$~g$^{-1}$ and that of a 1.764~MeV gamma-ray from $^{212}$Bi decay, for comparative purposes, is 0.0528~cm$^{2}$~g$^{-1}$. The flux ratio of the two isotopes can be found with the following equation:
\begin{align}\label{eq:eq2}
    \frac{I_{Tl}}{I_{Bi}} = e^{(A_{Bi} - A_{Tl})x}
\end{align}
where $x$ is distance in cm and $I$ is flux in cm$^{-2}$~s$^{-1}$. After 3.5~m of water shielding, the flux of the 2.615~MeV gamma-ray is 35 times higher. The 2.448~MeV line from $^{214}$Bi ($^{238}$U decay chain) also has a higher attenuation coefficient than the 2.615~MeV line, but this peak overlaps with the Q-value of 0νββ decay of $^{136}$Xe so the $^{238}$U background is still important for this search. The branching ratio of the line is low, at about 1.5\,\%, but a high concentration of $^{238}$U in rock can make this background dominant.

\autoref{tab:Flux} presents the flux of 2.615~MeV gamma-rays at each surface throughout the water shielding and the GdLS; the fluxes have been normalised to the first surface. The bases of each surface are closer together than each of their tops and sides, and there is a steel plate on the bottom of the water tank. Accounting for the impacts of these differences would be difficult and not worthwhile, thus, to calculate the attenuation coefficient, only the flux on the top and sides of each surface have been used. The attenuation coefficient decreases as distance increases because the gamma-rays with momentum in the direction of the TPC are the most likely to reach it as they will cross the least amount of water. At the first surface, the range of the gamma-rays' momentum vector is broad, but after several stages of re-propagation, the majority of gamma-rays are heading straight towards the detector.

\begin{table}[ht!]
\centering
\footnotesize
\def\arraystretch{1.1}%
\caption{Attenuation of 2.615~MeV gamma-rays. The surfaces in the top row are shown in \autoref{fig:geom_b}. The flux at each surface has been normalised to the flux at the first surface. The distance is measured from one surface to the next, corresponding to the fourth column in \autoref{tab:surfaces}. We have only used the top and sides of the surfaces here because the gamma-rays are not uniformly spread across the base of the water tank and detector, owing to the effect of the steel plate underneath. $m_i$ is the multiplication factor used in~\cref{eq:eq1} to boost gamma-rays from the current surface to the next, with each factor having a value that allowed $\sim10^7$ gamma-rays to reach the next surface. The flux ratio in the fifth row is the ratio between the flux at the current surface and the flux at the previous surface. The distances and the attenuation coefficients are cumulative, except for surfaces 5 to 6 where attenuation through the 0.5~m of GdLS is calculated separately, attributable to differences in material composition.}\label{tab:Flux}
\begin{tabular}{|c|c|c|c|c|cV{4}c|}
\hline
\multicolumn{1}{|c|}{\textbf{Surface}} & \multicolumn{1}{c|}{\textbf{1}} & \multicolumn{1}{c|}{\textbf{2}} & \multicolumn{1}{c|}{\textbf{3}} & \multicolumn{1}{c|}{\textbf{4}} & \multicolumn{1}{cV{4}}{\textbf{5}} & \multicolumn{1}{c|}{\textbf{6}}\\ 
\hline
\multicolumn{1}{|c|}{\textbf{Normalised flux}} & \multicolumn{1}{c|}{1.00} & \multicolumn{1}{c|}{$4.03\times10^{-3}$} & \multicolumn{1}{c|}{$3.82\times10^{-5}$} & \multicolumn{1}{c|}{$1.34\times10^{-6}$} & \multicolumn{1}{cV{4}}{$5.18\times10^{-8}$} & \multicolumn{1}{c|}{$8.36\times10^{-9}$} \\
\hline
\multicolumn{1}{|c|}{\textbf{Distance [cm]}} & \multicolumn{1}{c|}{0} & \multicolumn{1}{c|}{100} & \multicolumn{1}{c|}{200} & \multicolumn{1}{c|}{275} & \multicolumn{1}{cV{4}}{349} & \multicolumn{1}{c|}{50}\\
\hline
\multicolumn{1}{|c|}{\boldsymbol{$m_i$}} & \multicolumn{1}{c|}{110} & \multicolumn{1}{c|}{60} & \multicolumn{1}{c|}{10} & \multicolumn{1}{c|}{12} & \multicolumn{1}{cV{4}}{100} & \multicolumn{1}{c|}{5}\\
\hline
\multicolumn{1}{|c|}{\textbf{Flux ratio}} & \multicolumn{1}{c|}{n/a} & \multicolumn{1}{c|}{$4.03\times10^{-3}$} & \multicolumn{1}{c|}{$9.48\times10^{-3}$} & \multicolumn{1}{c|}{$3.50\times10^{-2}$} & \multicolumn{1}{cV{4}}{$3.87\times10^{-2}$} & \multicolumn{1}{c|}{0.161}\\
\hline
\multicolumn{1}{|c|}{\textbf{Attenuation}} & \multirow{2}{*}{n/a} & \multirow{2}{*}{$5.51\times10^{-2}$} & \multirow{2}{*}{$5.09\times10^{-2}$} & \multirow{2}{*}{$4.92\times10^{-2}$} & \multirow{2}{*}{$4.82\times10^{-2}$} & \multirow{2}{*}{$4.24\times10^{-2}$}\vspace{-1mm}\\
\multicolumn{1}{|c|}{\textbf{coefficient [cm$^{2}$~g$^{-1}$]}} & \multicolumn{1}{c|}{} & \multicolumn{1}{c|}{} & \multicolumn{1}{c|}{} & \multicolumn{1}{c|}{} & \multicolumn{1}{cV{4}}{} & \multicolumn{1}{c|}{}\\ 
\hline
\end{tabular}
\end{table}
The final stage of the simulation involves propagating the gamma-rays once more from the final, 6th surface. Information is collected from every particle that deposits more than 0.1~keV in the TPC, skin and GdLS volumes. These particles are primarily electrons with occasional positrons resulting from pair production. The information recorded includes energy deposits and their positions and time, as well as the parameters of the particles that release this energy in the detector volumes.

\subsection{Analysis and Results}
\label{sec:analysis}
A signal from either a WIMP interaction or a 0νββ decay will be a single energy deposition from one event. The number of events that would contribute to the background by depositing energy, $E_{dep}$, once in either the WIMP energy range or around the 0νββ decay signal, is extracted in the following analysis. The WIMP ROI in this analysis was 0~-~20~keV, similar to an extended ROI used to search for a wide range of WIMP masses~\cite{LZfirstResults}. For 0νββ decay, the energy resolution ($\sigma$/E) of the detector at 2.458~MeV was assumed to be 1\,\% from the LZ simulation model~\cite{LZ_0vbb}. The ROI considered in this analysis is conservatively taken as $\pm~2~\sigma$, 2.408~$<~E_{dep}~<$~2.508~MeV. It should be noted, however, that the energy resolution can be improved, as LZ has already demonstrated~\cite{energy_resolutionLZ}, and the energy ROI reduced to optimise the sensitivity. 

\begin{figure}[ht!] \centering
    \begin{subfigure}{.49\textwidth}
    \centering    \centering\includegraphics[width=1\textwidth,angle=0]{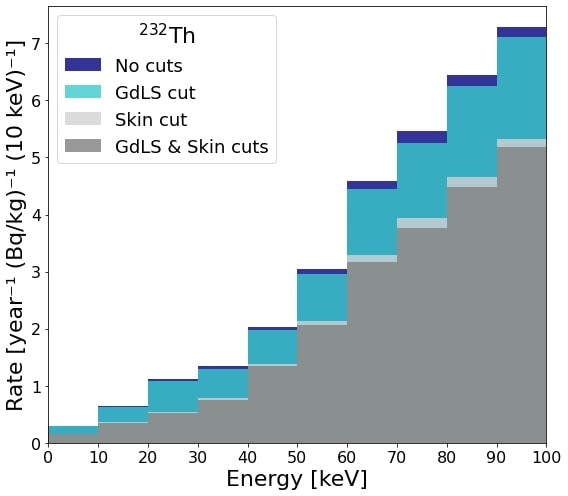}
        \caption{}
    \label{fig:cuts_th_100}
    \end{subfigure}
    \begin{subfigure}{.49\textwidth}
    \centering    \centering\includegraphics[width=1\textwidth,angle=0]{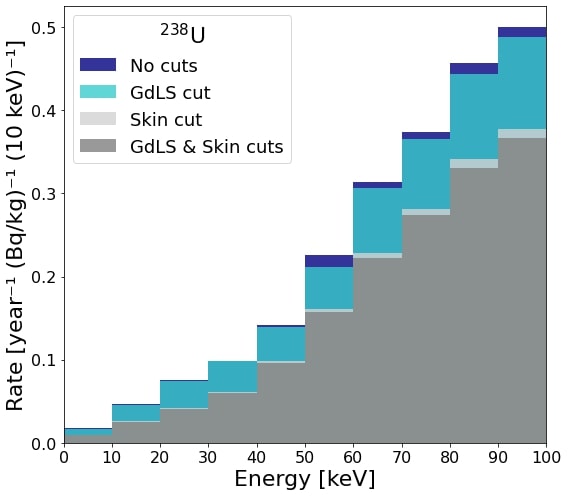}
        \caption{}
    \label{fig:cuts_u_100}
    \end{subfigure}
    \begin{subfigure}{.49\textwidth}
    \centering
    \centering\includegraphics[width=1\textwidth,angle=0]{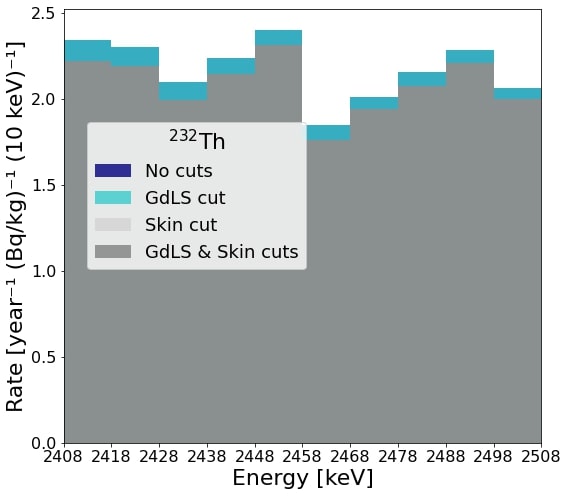}
        \caption{}
    \label{fig:cuts_th_Q}
    \end{subfigure}
    \begin{subfigure}{.49\textwidth}
    \centering
    \centering\includegraphics[width=1\textwidth,angle=0]{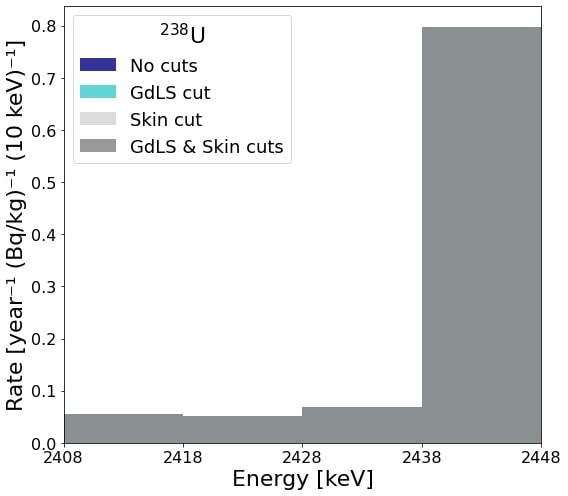}
        \caption{}
    \label{fig:cuts_u_Q}
    \end{subfigure}
    \caption{\small{Rate of energy deposits from the $^{232}$Th and $^{238}$U decay chains with the following cuts applied: the 200~keV GdLS threshold, the 100~keV skin threshold and both thresholds. The number of events from simulations before cuts at 0~-~20~keV is $2490\pm50$ in (a) and $576\pm24$ in (b), giving statistical errors of $\sim$\,2~\% and $\sim$\,4~\%, respectively. The average number of events at $\pm~50$~keV around the Q-value of $0\nu\beta\beta$ decay (2.458~MeV) in each bin is approximately 5700 in (c) and 2200 in (d).}}
    \label{fig:TPC_cuts}
\end{figure}

Core analysis cuts need to be implemented to suppress the gamma-ray background. The numbers of gamma-rays reaching the last surface prior to analysis cuts, taking into account the boosting factors, are equivalent to $9.63\times10^{17}$, $2.81\times10^{18}$ and $1.03\times10^{18}$ of initially generated gamma-rays in rock for $^{232}$Th, $^{238}$U, and $^{40}$K, respectively. The baseline threshold for the GdLS was chosen to be 200~keV to avoid false vetoes from the decays of $^{14}$C, $^{152}$Gd, and $^{147}$Sm~\cite{LZ_sim}. A similar threshold for events in the skin was taken as 100~keV, which is sufficient to detect most Compton electrons from MeV gamma-rays~\cite{LZ_wimp_sensitivity,LZ_tdr}. To reduce the amount of data stored on disk, information about energy depositions in the GdLS and skin was only saved at this last stage of the simulation where the gamma-rays are transported from the 6th surface to the TPC. Any depositions from previous stages have not been stored and thus, the results presented here are conservative. In the analysis of real data, gamma-rays may leave a signal in the GdLS before reaching the TPC. These cuts were implemented by adding up all energy depositions of an event in the TPC within 1~$\mu$s (anti-coincidence time window), then similarly and separately adding up all energy depositions of the same event in the skin and scintillator, also within 1~$\mu$s. If an event in the TPC had a companion event in either the skin or liquid scintillator, within 1~$\mu$s and above their respective thresholds, this event is rejected. An example of the effect of these cuts applied to raw data is displayed in \autoref{fig:TPC_cuts}.

\begin{figure}[ht!] 
\centering
\begin{subfigure}{0.325\textwidth}
\includegraphics[width=\linewidth,angle=0]{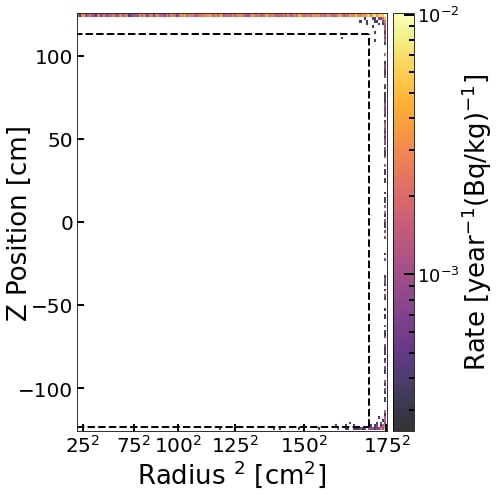}
\setcounter{subfigure}{0}
\vspace{-8mm}
\caption{} \label{fig:MS_cut_20a}
\end{subfigure}
\begin{subfigure}{0.325\textwidth}
\includegraphics[width=\linewidth,angle=0]{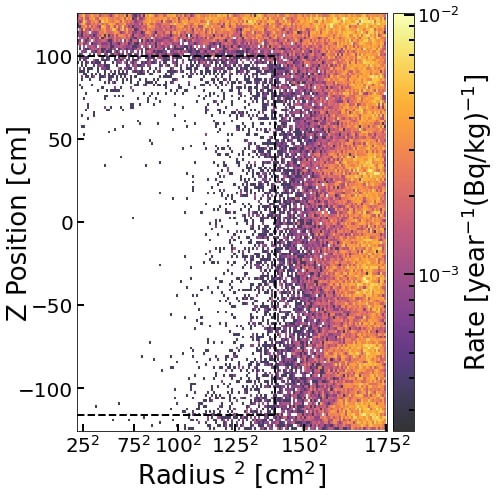}
\setcounter{subfigure}{2}
\vspace{-8mm}
\caption{} \label{fig:MS_cut_Qc}
\end{subfigure}
\begin{subfigure}{0.325\textwidth}
\includegraphics[width=\linewidth,angle=0]{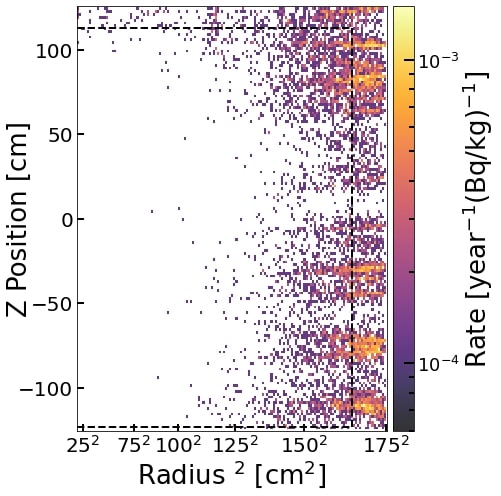}
\setcounter{subfigure}{4}
\vspace{-8mm}
\caption{} \label{fig:MS_cut_Qe}
\end{subfigure}
\vspace{-3mm}
\medskip
\begin{subfigure}{0.325\textwidth}
\includegraphics[width=\linewidth,angle=0]{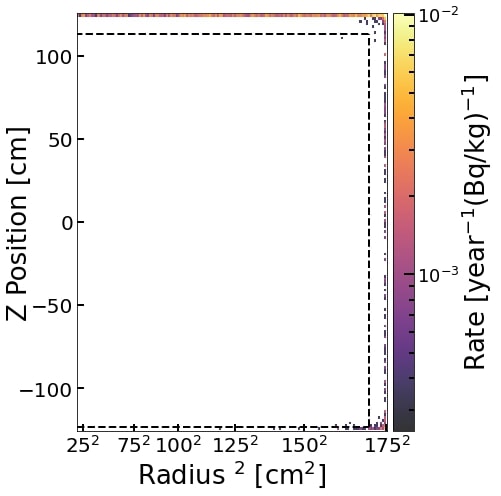}
\setcounter{subfigure}{1}
\vspace{-8mm}
\caption{} \label{fig:MS_cut_20b}
\end{subfigure}
\begin{subfigure}{0.325\textwidth}
\includegraphics[width=\linewidth,angle=0]{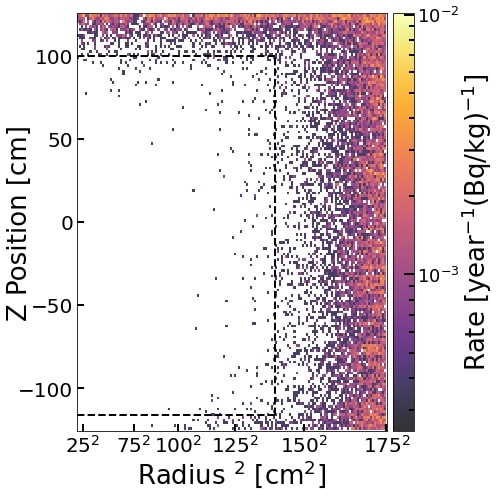}
\setcounter{subfigure}{3}
\vspace{-8mm}
\caption{} \label{fig:MS_cut_Qd}
\end{subfigure}
\begin{subfigure}{0.325\textwidth}
\includegraphics[width=\linewidth,angle=0]{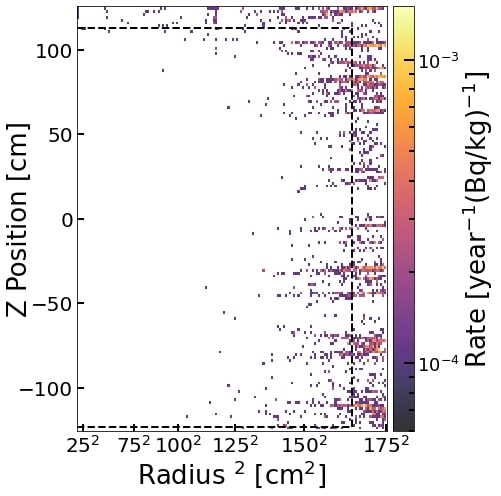}
\setcounter{subfigure}{5}
\vspace{-8mm}
\caption{} \label{fig:MS_cut_Qf}
\end{subfigure}
    \caption{Scatter plots of the mean position of events from the $^{232}$Th and $^{238}$U decay chains in the TPC.~\cref{fig:MS_cut_20a,fig:MS_cut_20b} show $^{232}$Th chain event positions within the energy range 0~-~20~keV (WIMP region of interest) before and after multiple scatter cuts have been applied, respectively. The dotted black line outlines the fiducial volume boundaries for WIMP search. Similarly, in~\cref{fig:MS_cut_Qc,fig:MS_cut_Qd}, the mean position of events from $^{232}$Th in the energy range 2.408~-~2.508~MeV (50~keV around the 0νββ decay Q-value) are depicted before and after multiple scatter cuts have been applied, respectively. The dotted black line outlines the reduced fiducial volume that will need to be in place for 0νββ decay background sensitivity.~\cref{fig:MS_cut_Qe,fig:MS_cut_Qf} are the equivalent to~\ref{fig:MS_cut_Qc} and~\ref{fig:MS_cut_Qd}, but for $^{238}$U chain events. Plots of $^{238}$U events in the WIMP ROI are not shown here as they are similar to those from $^{232}$Th shown in 5a and 5b. The clusters of events apparent in~\cref{fig:MS_cut_Qe,fig:MS_cut_Qf} result from boosting, exacerbated by the fact that a high-energy (2.448 MeV) gamma-ray from the uranium chain must deposit almost all of its energy in a single interaction to be included in the 0$\nu\beta\beta$ ROI.}
    \label{fig:MS_cut_Q}
\end{figure}

The $x$-, $y$-, and $z$-positions of energy depositions and the amount of energy released by each energy deposition are stored. The $z$-axis spans the entire height of the TPC, while the orthogonal $x$- and $y$-axes, each corresponding to the TPC’s diameter, are combined to parameterise the radius, $r$, of the TPC. To reject multiple scattering gamma-rays, we require the energy-weighted standard deviations ($\sigma_z$ and $\sigma_r$) of separate clusters of energy deposits per event to be less than the spatial resolution of the detector. $\sigma_z$ and $\sigma_r$ are calculated using~\cref{eq:eq3,eq:eq4}, respectively:
\begin{gather}\label{eq:eq3}
    \sigma_z=\sqrt{\frac{\sum^N_{i=1}w_i(z_i-\bar{z})^2}{\sum^N_{i=1}w_i}}
\end{gather}
\begin{gather}\label{eq:eq4}
    \sigma_r=\sqrt{\frac{\sum^N_{i=1}w_i((x_i-\bar{x})^2+(y_i-\bar{y})^2)}{\sum^N_{i=1}w_i}}
\end{gather}
where for an energy deposition, $i$, in an event with $N$ deposits, $w_i$ is the fraction of the event's total energy released by $i$, and $x_i$, $y_i$ and $z_i$ are the positions of $i$ in the TPC. All positions in an event have been averaged and weighted with $w$ to obtain $\bar{x}$, $\bar{y}$ and $\bar{z}$, the weighted mean positions of energy deposits.
Similar work has been done by LZ~\cite{LZ_sim} and LUX~\cite{LUX,Pos_reconst}. Based on position reconstruction carried out by LUX and accounting for the up-scaled size of the detector, the spatial resolutions of the future detector were conservatively assumed to be 0.5~cm in the vertical direction and 5~cm in the horizontal direction. Thus, it was assumed that events with $\sigma_z>$~0.5~cm and $\sigma_r>$~5.0~cm will be reconstructed as multiple scatter events and will be rejected. These thresholds are larger than those used for the LUX and LZ experiments because the increased size of the detector causes a decrease in resolution. 
Lower $\sigma_z$ and $\sigma_r$ thresholds would contribute to further improvements in sensitivity and a reduction in background rates for WIMP and 0$\nu\beta\beta$ decay search. The numbers of multiple scattered events for $^{232}$Th, $^{238}$U and $^{40}$K are shown in~\ref{sec:apx:std_dev}, and it is clear that at higher energies there are many more of these events. The impact of the multiple scatter cut on the rate in the TPC is therefore greater at higher energies, as is demonstrated in \autoref{fig:MS_cut_Q}.

Lastly, a fiducial volume (FV) cut vetoes events near the edges of the TPC where the bulk of gamma-ray radiation, from the rock and also detector components, appears. The attenuation lengths of low-energy ($\sim10^2$~keV) gamma-rays in LXe are around a few centimetres~\cite{NIST}, thus to remove these background events for the WIMP ROI we have set a cylindrical ``WIMP FV'' with boundaries: -123~cm~$<z<113$~cm and radius $<170$~cm, which translates to 2~cm from the cathode grid at the base of the TPC, 12~cm from the gate grid at the top of the TPC and 5~cm from the TPC wall. These boundaries, containing approximately 63.3~tonnes of LXe, are outlined in~\cref{fig:MS_cut_20a,fig:MS_cut_20b}, where the effect of the multiple scatters cut on events in the WIMP ROI can also be seen. Almost all events in the energy range 0~-~20~keV are single scatters and mainly located near the TPC walls outside of the WIMP FV. 

To remove background events from the 0νββ decay ROI the FV will need to be much smaller to account for the longer attenuation lengths of higher energy gamma-rays. The FV limits were minimised until the rate of events from $^{232}$Th and $^{238}$U were each $<0.1$~year$^{-1}$~(Bq/kg)$^{-1}$. These reduced ``0νββ FVs'' are illustrated in~\cref{fig:MS_cut_Qc,fig:MS_cut_Qf} and correspond to fiducial masses of 39.3~tonnes and 59.6~tonnes for $^{232}$Th and $^{238}$U, respectively. The boundaries of these 0νββ FVs are: 9~cm and 2~cm from the cathode grid; 25~cm and 12~cm from the gate grid; and 35~cm and 10~cm from the TPC wall (the first value corresponds to the background from $^{232}$Th and the second value from $^{238}$U). The most stringent FV of the two will be chosen for the experiment, and this will be determined by the concentration of $^{232}$Th or $^{238}$U in the cavern rock around the experiment. Note that the FV is expected to be smaller than anticipated here due to additional, and higher, contributions from the radioactivity in detector components not considered here. Further discussion on reducing the FV for 0νββ decay search is in Section~\ref{sec:discussion}. The effects of all analysis cuts on the total number of events from each nuclide are summarised in \autoref{tab:cuts}.

\begin{table}[ht!]
    \centering
    \footnotesize
    \def\arraystretch{1.2}
    \setlength{\tabcolsep}{4pt}
        \caption{A summary of the analysis cuts outlined in Section~\ref{sec:analysis} and their effects on the number of events depositing energy in the TPC. After each cut, the remaining events are expressed as a fraction of the total number of events. GdLS and Skin thresholds are applied in the first row, then the $\sigma_r$ and $\sigma_z$ multiple scatter cuts in rows 2 and 3, respectively. The last row includes the fiducial volume cuts, displaying the final fraction of total events that survive all of the analysis cuts.}
    \begin{tabular}{|c|c|c|c|c|c|}
        \hline
        \multicolumn{1}{|c}{} &  \multicolumn{2}{|c}{\textbf{$^{232}$Th}} & \multicolumn{2}{|c}{\textbf{$^{238}$U}} & \multicolumn{1}{|c|}{\textbf{$^{40}$K}}\\
        \cline{2-6}
        \textbf{Cuts} & \textbf{0 - 100 keV} & \textbf{2408 - 2508 keV} & \textbf{0 - 100 keV} & \textbf{2408 - 2508 keV} & \textbf{0 - 100 keV}\\
        \hline
        GdLS \& Skin & 0.68 & 0.93 & 0.67 & 1.00 & 0.76 \\
        \hline
        MS $(x,y)$ & 0.66 & 0.75 & 0.67 & 0.80 & 0.74 \\
        \hline 
        MS $(z)$ & 0.66 & 0.27 & 0.66 & 0.29 & 0.74 \\
        \hline
        FV & 4.7$\times10^{-4}$ & 4.4$\times10^{-3}$ & 7.4$\times10^{-4}$ & 2.9$\times10^{-2}$ & 0 \\
        \hline
    \end{tabular}
    \label{tab:cuts}
\end{table}
The event rate for each radioactive nuclide depositing energy in the TPC per year, normalised to 1~Bq~kg$^{-1}$ is presented in \autoref{tab:Rates}. Asymmetric uncertainties are quoted at 68.27\,\% confidence level (C.L.) limits for the Poisson signal mean and 90\,\% C.L. limits for 0 values. The 0~-~100~keV rates are presented here because this covers an extended range of energies of interest for WIMP-nucleon effective field theory couplings~\cite{EFT}. The rates in the energy range 0~-~20~keV and 0~-~100~keV include the WIMP FV cut, and the rates in the 2408~-~2508~keV energy range include the 0νββ FV cut corresponding to either $^{232}$Th or $^{238}$U. Because the total statistics for $^{238}$U is larger than that of $^{232}$Th, and the 0νββ FV is also larger, the number of events in the 0νββ ROI is larger for $^{238}$U compared to $^{232}$Th, while the rates are similar. In addition to these analysis cuts, discrimination between NRs and ERs will contribute to the suppression of background events for WIMP search~\cite{LZfirstResults}. 

\begin{center}
\footnotesize
\def\arraystretch{1.3}
\setlength{\tabcolsep}{4pt}
\begin{longtable}[htbp]{|c|c|c|c|c|c|c|c|}
\caption{Background events in the TPC with the analysis cuts applied. The WIMP fiducial volume (FV) cut is applied for 0~–~20~keV and 0~–~100~keV energy ranges. For 2408~-~2508~keV, the $^{232}$Th 0νββ FV cut is applied to $^{232}$Th rates, while the $^{238}$U 0νββ FV cut is applied to $^{238}$U rates. See Section~\ref{sec:analysis} text for the evaluation of the WIMP and 0νββ FVs. Using a Feldman-Cousins approach~\cite{FC_stats}, the statistical uncertainties are shown at 1~$\sigma$ for non-zero event rates and an upper limit of 90\,\% C.L. is quoted for zero events.}\label{tab:Rates} \\
\hline
\multicolumn{1}{|c|}{} & \multicolumn{2}{c|}{\textbf{0 - 20~keV}} & \multicolumn{2}{c|}{\textbf{0 - 100~keV}} & \multicolumn{2}{c|}{\textbf{2408 - 2508 keV}} \\
\hline
\multicolumn{1}{|c|}{\textbf{Isotope}} & \multicolumn{1}{c|}{\textbf{Events}} & \multicolumn{1}{c|}{\textbf{Rate [year$^{-1}$}} & \multicolumn{1}{c|}{\textbf{Events}} & \multicolumn{1}{c|}{\textbf{Rate [year$^{-1}$}} & \multicolumn{1}{c|}{\textbf{Events}} & \multicolumn{1}{c|}{\textbf{Rate [year$^{-1}$}} \\ [-0.22cm]
\multicolumn{1}{|c|}{} & \multicolumn{1}{c|}{} & \multicolumn{1}{c|}{\textbf{(Bq/kg)$^{-1}$]}} & \multicolumn{1}{c|}{} & \multicolumn{1}{c|}{\textbf{(Bq/kg)$^{-1}$]}} & \multicolumn{1}{c|}{} & \multicolumn{1}{c|}{\textbf{(Bq/kg)$^{-1}$]}} \\
\hline
\endfirsthead \hline
\endlastfoot
$^{232}$Th & $4_{-1.66}^{+2.78}$ & $\left(1.5_{-0.6}^{+1.1}\right)\times10^{-3}$ & $37\pm6.1$ & $\left(1.42\pm0.23\right)\times10^{-2}$ & $253\pm 15.9$ & $\left(9.69\pm0.61\right)\times10^{-2}$ \\
\hline
$^{238}$U & $2_{-1.26}^{+2.25}$ & $\left(2.2_{-1.4}^{+2.5}\right)\times10^{-4}$ & $14_{-3.70}^{+4.32}$ & $\left(1.6_{-0.4}^{+0.5}\right)\times10^{-3}$ & $820\pm 28.6$ & $\left(9.19\pm0.32\right)\times10^{-2}$ \\
\hline
$^{40}$K & $0_{-0}^{+2.44}$ & $\left(0_{-0}^{+3.8}\right)\times10^{-5}$ & $0_{-0}^{+2.44}$ & $\left(0_{-0}^{+3.8}\right)\times10^{-5}$ & n/a & n/a \\
\hline
\end{longtable}
\end{center}

\section{Measurements of gamma-rays from rock samples in the Boulby Underground Laboratory}
\label{sec:boulby}
The majority of rock samples were collected by the ICL-Boulby exploratory geology team via boreholes which span the various strata of rock types that are vertically spread over roughly 60~m from $\sim\,$1040~to~1100~m below the surface. The mine's rock is predominantly halite-based as the area used to be part of the Zechstein Sea millions of years ago. Around a depth of 1040~m, the rock tends to be mainly potash, clear pink halite (CPH) and footwall halite (FWHL). Venturing deeper, there are bands of different types of halite and salt polygons, before reaching anhydrite. It should be noted that the stratum of each rock type undulates across hundreds of metres horizontally around the 1100~m level~\cite{rock_layers}. Samples of polyhalite were collected from different boreholes to those mentioned previously, at depths of around 1100~m and 1300~m.

\subsection{Methods for sample measurement}
\label{measurements}
The collected rock samples were placed individually in a sample holder within a grip seal plastic bag, then positioned inside ``Chaloner'' and were measured for 117~-~740~hours. Chaloner is a Mirion BE5030 broad-energy, HPGe detector, which can measure gamma-rays in the range 3~keV~-~3~MeV with a relative efficiency of 48\,\%~\cite{BUGS}. The sample holder tube is made from a mixture of acrylic and PMMA while the base disc material is perspex. LabSOCS (Laboratory Sourceless Calibration Software)~\cite{LabSOCS} can simulate each of the germanium detectors and any sample placed on it within a sample holder. Depending on the density, size, geometry and alignment of the sample, LabSOCS will generate an efficiency curve spanning the required energy range. An example of a spectrum produced by Chaloner measuring a sample of anhydrite can be seen in~\ref{sec:apx:sample}. The software InterSpec~\cite{InterSpec,InterSpec2} automatically fitted each peak with a Gaussian. The number of gamma-rays from each fitted Gaussian is recorded and normalised with respect to the sample's mass, the energy-specific detector efficiency, the time the sample was measured for and the branching ratio of the radioactive nuclide that produced the gamma-rays of the peak. The subtracted background is that of the empty sample holder in a plastic bag.

\subsection{Results of sample measurement}
\label{sec:boulby_results}
\autoref{tab:samples} lists the activities of $^{40}$K, $^{232}$Th, $^{238}$U and $^{235}$U in 10 types of rock. Three samples of halite are included to emphasise the variety in the abundance of radioactive contaminations in the salt band at the 1100~m level. Although all samples from the 1100~m level are mainly salt rock in composition (halites), some have more contaminants than others and consequently have differing abundances of radioactive nuclides. \autoref{tab:samples} includes the measurements of $^{40}$K, $^{232}$Th, $^{238}$U and $^{235}$U in 10 types of rock from mainly the 1100~m level, but also some polyhalite from the 1300~m level. Measurements of multiple samples of each rock type were averaged. Some of the $^{235}$U measurements have upper limits quoted at 95\,\% C.L. because, in the gamma-ray energy spectra for these samples, the uncertainties of the peak counts were very high. This occurred when the potassium content tended to be very large and the $^{235}$U content relatively low. The resulting, extensive Compton continuum obscured multiple smaller peaks below the $^{40}$K peak at 1461~keV. 

\begin{center}
\scriptsize
\def\arraystretch{1.5}%
\setlength{\tabcolsep}{4pt}
\begin{longtable}[hbt!]{|c|c|c|c|c|}
\caption{Averaged measurements of radioactive isotopes in rock samples from Boulby mine. Statistical uncertainties as a standard deviation are quoted first and systematic uncertainties second. 
Where measurements of $^{235}$U were statistically insignificant, an upper limit at 95\,\% C.L. is reported. Systematic uncertainties are outlined in~\ref{sec:apx:boulby_sys_uncert}.} \label{tab:samples} \\
\hline
\multicolumn{1}{|c|}{\fontsize{8.5}{9.5}\selectfont\textbf{Rock type}} & \multicolumn{1}{c|}{\fontsize{8.5}{9.5}\selectfont\textbf{$^{40}$K activity}} & \multicolumn{1}{c|}{\fontsize{8.5}{9.5}\selectfont\textbf{$^{232}$Th activity}} & \multicolumn{1}{c|}{\fontsize{8.5}{9.5}\selectfont\textbf{$^{238}$U activity}} & \multicolumn{1}{c|}{\fontsize{8.5}{9.5}\selectfont\textbf{$^{235}$U activity}} \\ 
\multicolumn{1}{|c|}{} & \multicolumn{1}{c|}{\fontsize{8.5}{9.5}\selectfont\textbf{[Bq kg$^{-1}$]}} & \multicolumn{1}{c|}{\fontsize{8.5}{9.5}\selectfont\textbf{[Bq kg$^{-1}$]}} & \multicolumn{1}{c|}{\fontsize{8.5}{9.5}\selectfont\textbf{[Bq kg$^{-1}$]}} & \multicolumn{1}{c|}{\fontsize{8.5}{9.5}\selectfont\textbf{[Bq kg$^{-1}$]}} \\ \hline 
\endfirsthead \hline
\endlastfoot
$\fontsize{8.5}{9.5}\selectfont\text{Polyhalite}$ & {\fontsize{8.5}{9.5}\selectfont $2500\pm 1\pm 126$} & {\fontsize{8.5}{9.5}\selectfont $\left(1.14 \pm 0.37 \pm0.06\right)\times10^{-2}$} & {\fontsize{8.5}{9.5}\selectfont $0.354 \pm 0.008 \pm 0.042$} & {\fontsize{8.5}{9.5}\selectfont $<3.4\times10^{-2}$} \\
\hline
{\fontsize{8.5}{9.5}\selectfont Salt polygons} & {\fontsize{8.5}{9.5}\selectfont $47.09\pm0.27\pm2.66$} & {\fontsize{8.5}{9.5}\selectfont $0.161\pm0.005\pm0.004$} & {\fontsize{8.5}{9.5}\selectfont $0.145\pm0.006\pm0.004$} & {\fontsize{8.5}{9.5}\selectfont $<1.3\times10^{-2}$ }\\
\hline
{\fontsize{8.5}{9.5}\selectfont LG potash} & {\fontsize{8.5}{9.5}\selectfont $3578 \pm 3\pm88$} & {\fontsize{8.5}{9.5}\selectfont $3.364\pm0.028\pm0.096$} & {\fontsize{8.5}{9.5}\selectfont $2.514\pm0.029\pm0.294$} & {\fontsize{8.5}{9.5}\selectfont $0.139\pm0.009\pm0.008$} \\
\hline
{\fontsize{8.5}{9.5}\selectfont Potash} & {\fontsize{8.5}{9.5}\selectfont $1508 \pm 3\pm43$} & {\fontsize{8.5}{9.5}\selectfont $2.756\pm0.041\pm0.322$} & {\fontsize{8.5}{9.5}\selectfont $2.279
\pm0.041\pm0.128$} & {\fontsize{8.5}{9.5}\selectfont $0.117\pm0.011\pm0.004$ }\\
\hline
{\fontsize{8.5}{9.5}\selectfont FWHL} & {\fontsize{8.5}{9.5}\selectfont $281.9\pm1.2\pm32.9$} & {\fontsize{8.5}{9.5}\selectfont $1.144\pm0.019\pm0.064$} & {\fontsize{8.5}{9.5}\selectfont $1.140\pm0.021\pm0.042$} & {\fontsize{8.5}{9.5}\selectfont $\left(6.02\pm0.48\pm0.28\right)\times10^{-2}$} \\
\hline
{\fontsize{8.5}{9.5}\selectfont CPH} & {\fontsize{8.5}{9.5}\selectfont $1709\pm3\pm96$} & {\fontsize{8.5}{9.5}\selectfont $0.358\pm0.023\pm0.013$} & {\fontsize{8.5}{9.5}\selectfont $0.492\pm0.025\pm0.023$} & {\fontsize{8.5}{9.5}\selectfont $<4.2\times10^{-2}$} \\
\hline
{\fontsize{8.5}{9.5}\selectfont Anhydrite} & {\fontsize{8.5}{9.5}\selectfont $13.93\pm0.17\pm0.51$} & {\fontsize{8.5}{9.5}\selectfont $0.610\pm0.008\pm0.028$} & {\fontsize{8.5}{9.5}\selectfont $2.919\pm0.017\pm0.162$} & {\fontsize{8.5}{9.5}\selectfont $0.139\pm0.003\pm0.009$ }\\
\hline
{\fontsize{8.5}{9.5}\selectfont Halite 3} & {\fontsize{8.5}{9.5}\selectfont $584.4\pm1.7\pm27.3$} & {\fontsize{8.5}{9.5}\selectfont $0.888\pm0.022\pm0.049$} & {\fontsize{8.5}{9.5}\selectfont $0.826\pm0.022\pm0.032$} & {\fontsize{8.5}{9.5}\selectfont $<5.1\times10^{-2}$ } \\
\hline
{\fontsize{8.5}{9.5}\selectfont Halite 4} & {\fontsize{8.5}{9.5}\selectfont $479.9\pm0.9\pm26.7$} & {\fontsize{8.5}{9.5}\selectfont $4.262\pm0.020\pm0.164$} & {\fontsize{8.5}{9.5}\selectfont $2.323\pm0.018\pm0.107$} & {\fontsize{8.5}{9.5}\selectfont $0.129\pm0.004\pm0.011$ }\\
\hline 
{\fontsize{8.5}{9.5}\selectfont Halite 9} & {\fontsize{8.5}{9.5}\selectfont $39.06\pm0.14\pm1.51$} & {\fontsize{8.5}{9.5}\selectfont $0.529\pm0.005\pm0.024$} & {\fontsize{8.5}{9.5}\selectfont $0.495\pm0.005\pm0.029$} & {\fontsize{8.5}{9.5}\selectfont $\left(4.37\pm0.12\pm0.25\right)\times10^{-2}$ }\\
\hline
\end{longtable}
\end{center}

\begin{figure}[ht!] 
\centering
\begin{subfigure}{0.32\textwidth}
\includegraphics[width=\linewidth,angle=0]{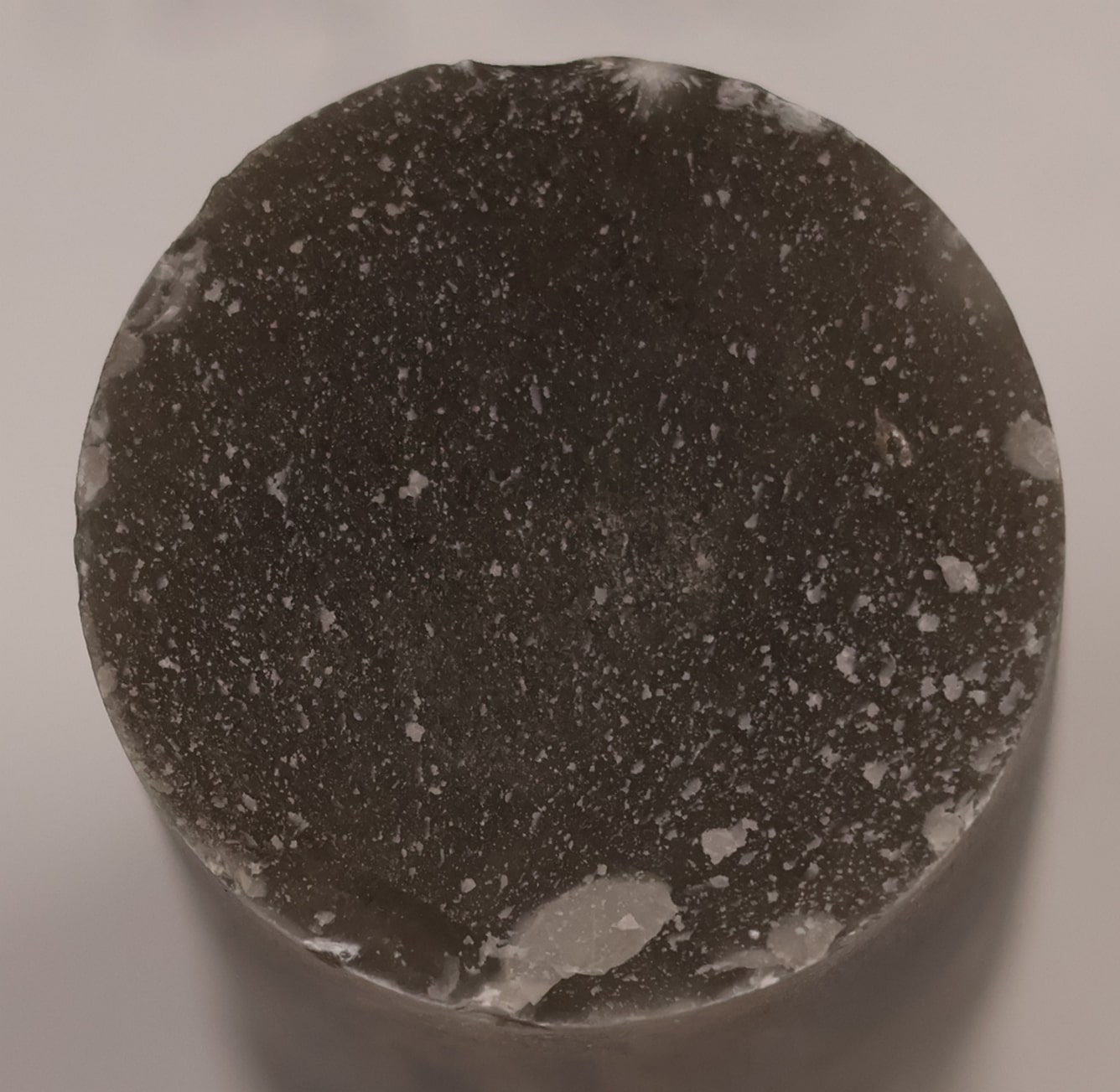}
\caption{Polyhalite, 1100~m depth}\label{fig:Rocks_a}
\end{subfigure}
\begin{subfigure}{0.32\textwidth}
\includegraphics[width=\linewidth,angle=0]{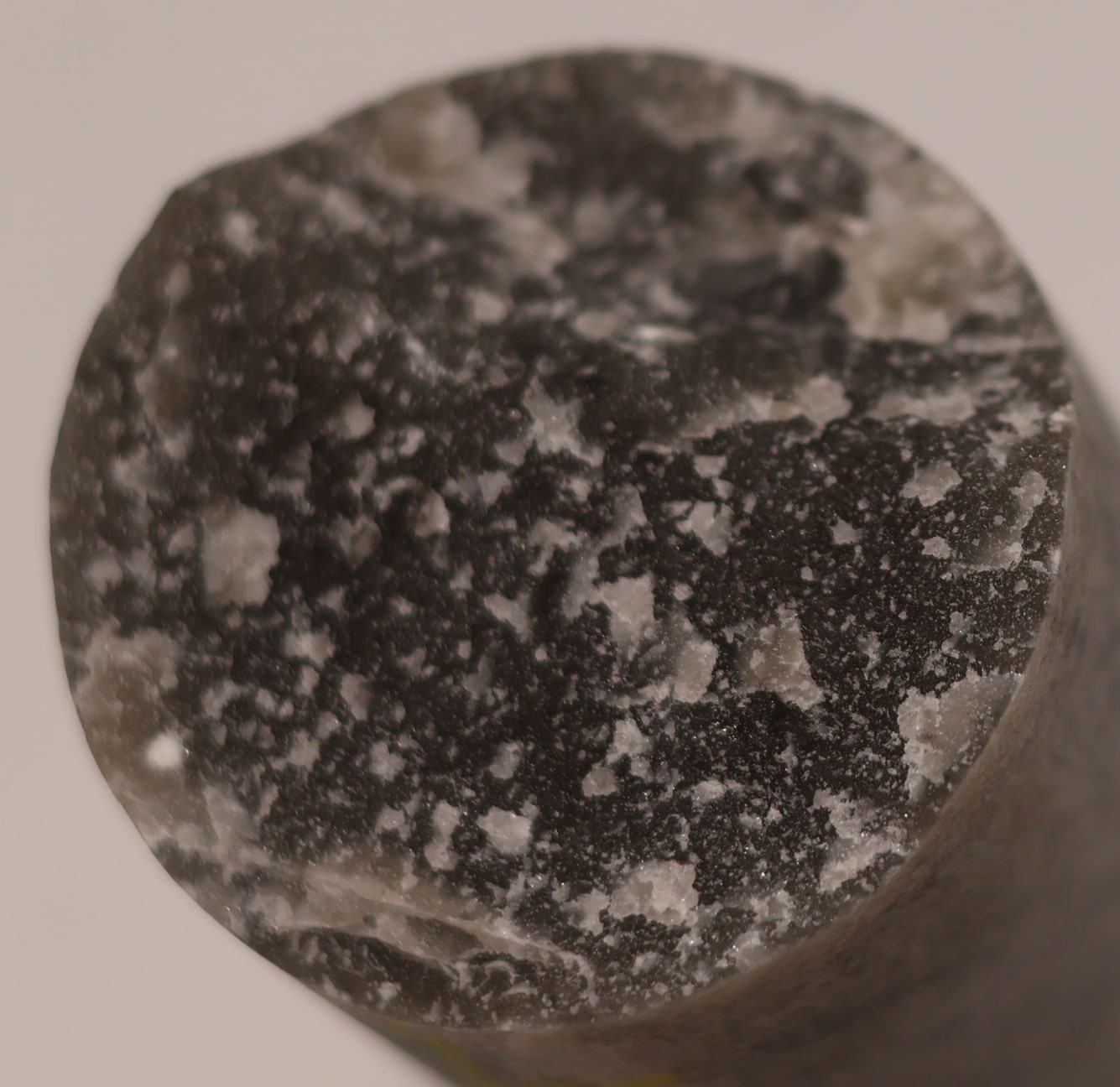}
\caption{Polyhalite, 1300~m depth} \label{fig:Rocks_h}
\end{subfigure}
\begin{subfigure}{0.32\textwidth}
\includegraphics[width=\linewidth,angle=0]{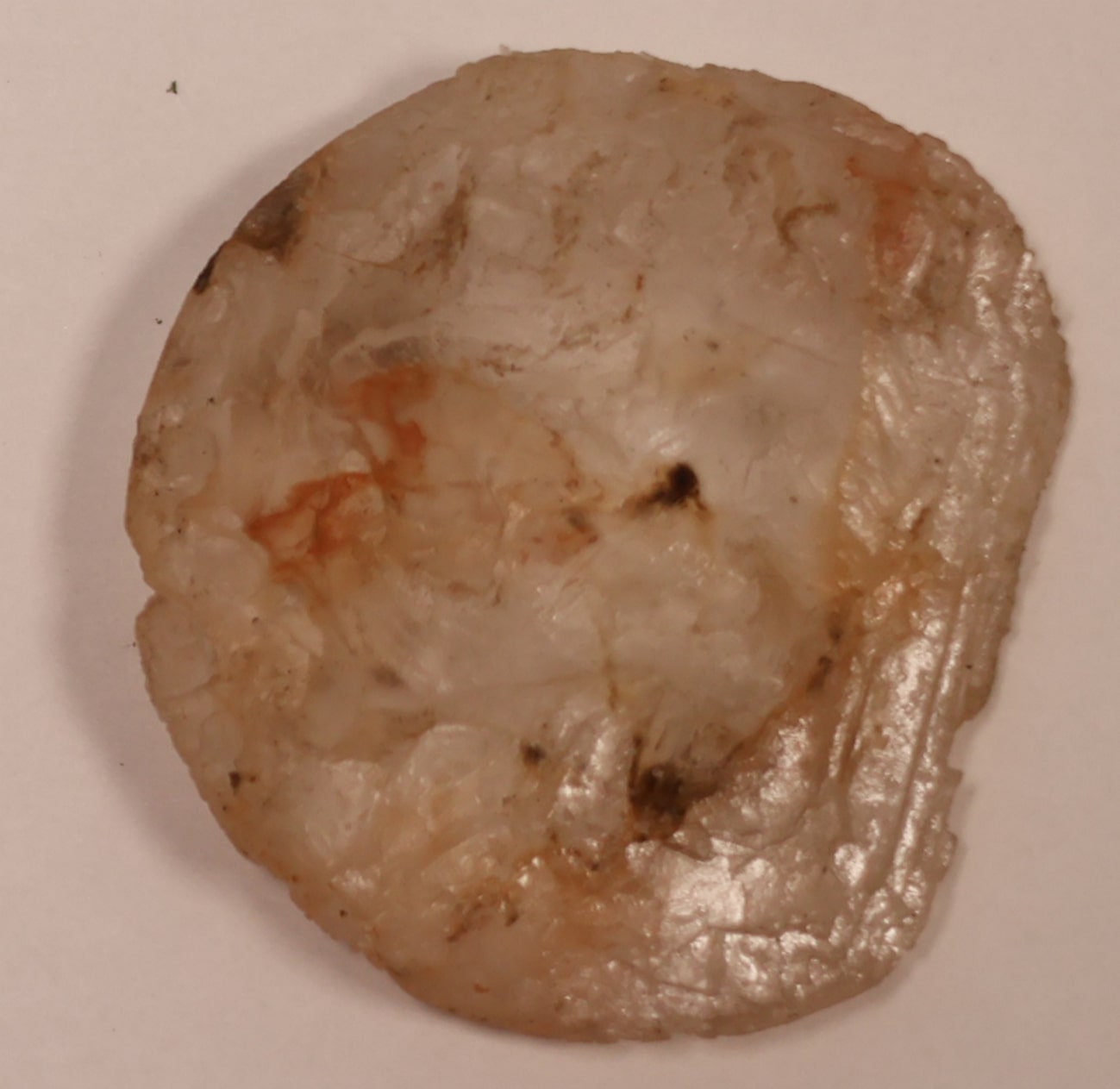}
\caption{Salt polygons}\label{fig:Rocks_b}
\end{subfigure}
\begin{subfigure}{0.32\textwidth}
\includegraphics[width=\linewidth,angle=0]{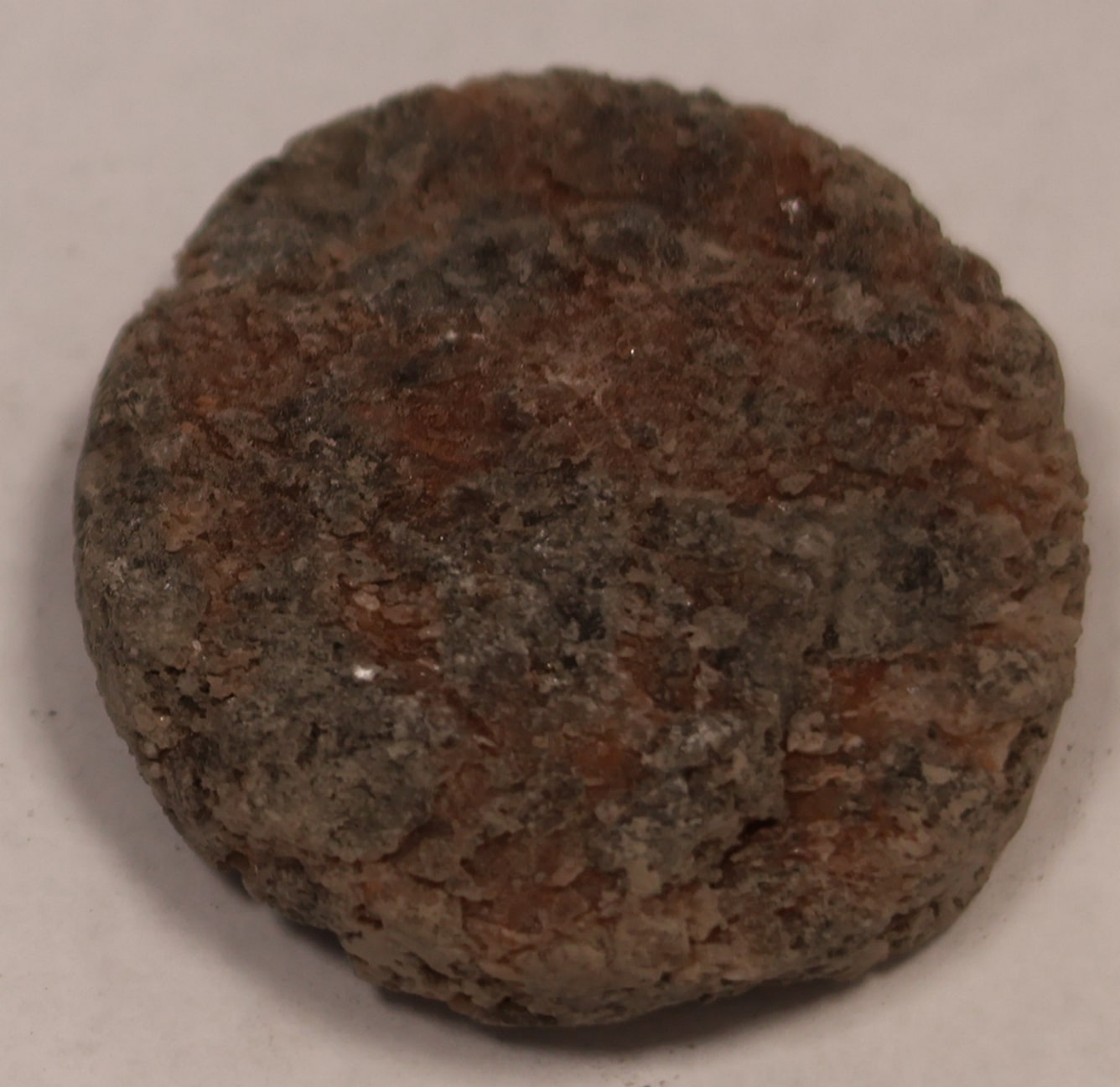}
\caption{LG potash}\label{fig:Rocks_c}
\end{subfigure}
\medskip
\begin{subfigure}{0.32\textwidth}
\includegraphics[width=\linewidth,angle=0]{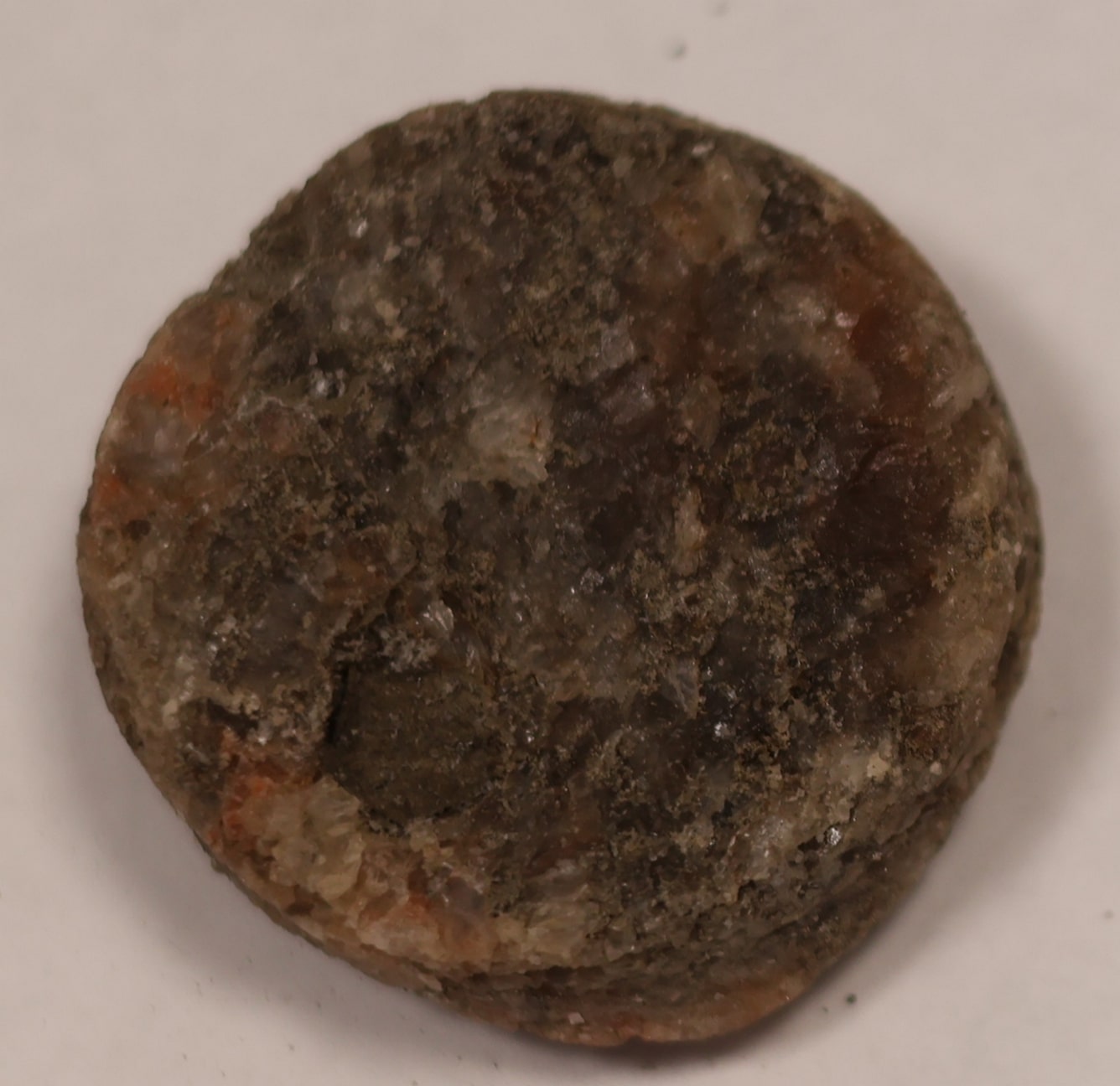}
\caption{FWHL} \label{fig:Rocks_d}
\end{subfigure}
\begin{subfigure}{0.32\textwidth}
\includegraphics[width=\linewidth,angle=0]{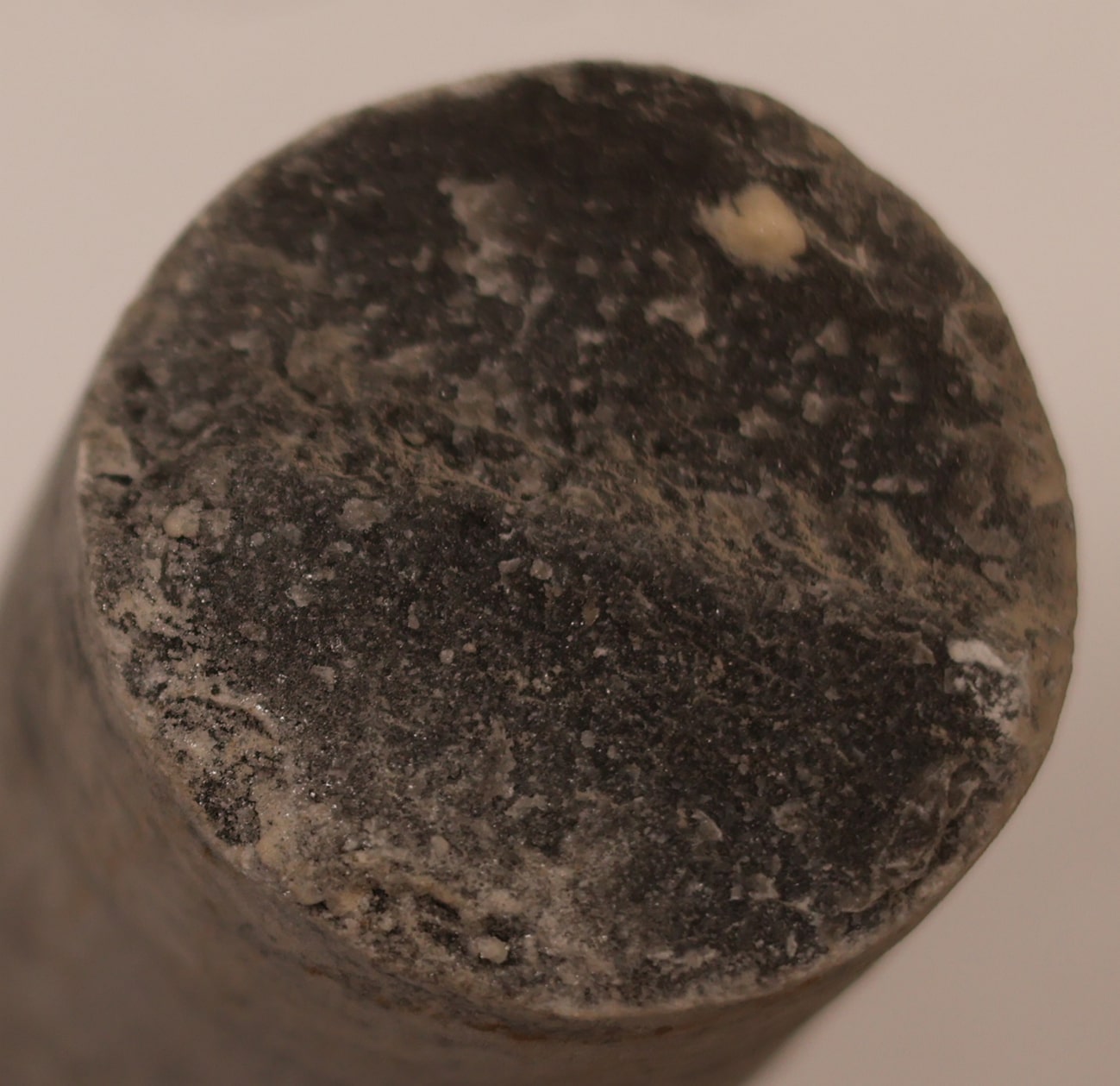}
\caption{Anhydrite} \label{fig:Rocks_e}
\end{subfigure}
\begin{subfigure}{0.32\textwidth}
\includegraphics[width=\linewidth,angle=0]{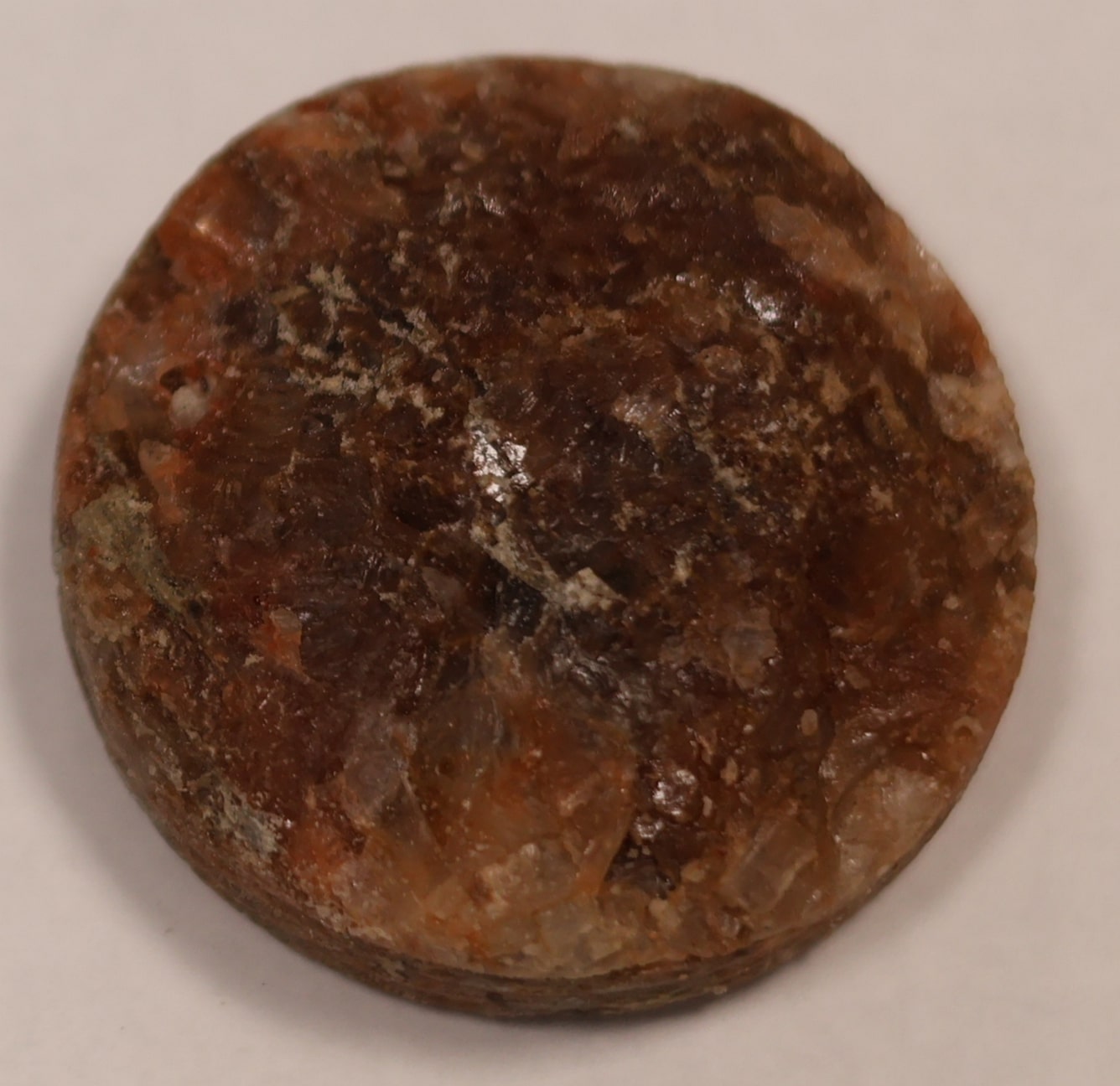}
\caption{Halite 3} \label{fig:Rocks_f}
\end{subfigure}
\begin{subfigure}{0.32\textwidth}
\includegraphics[width=\linewidth,angle=0]{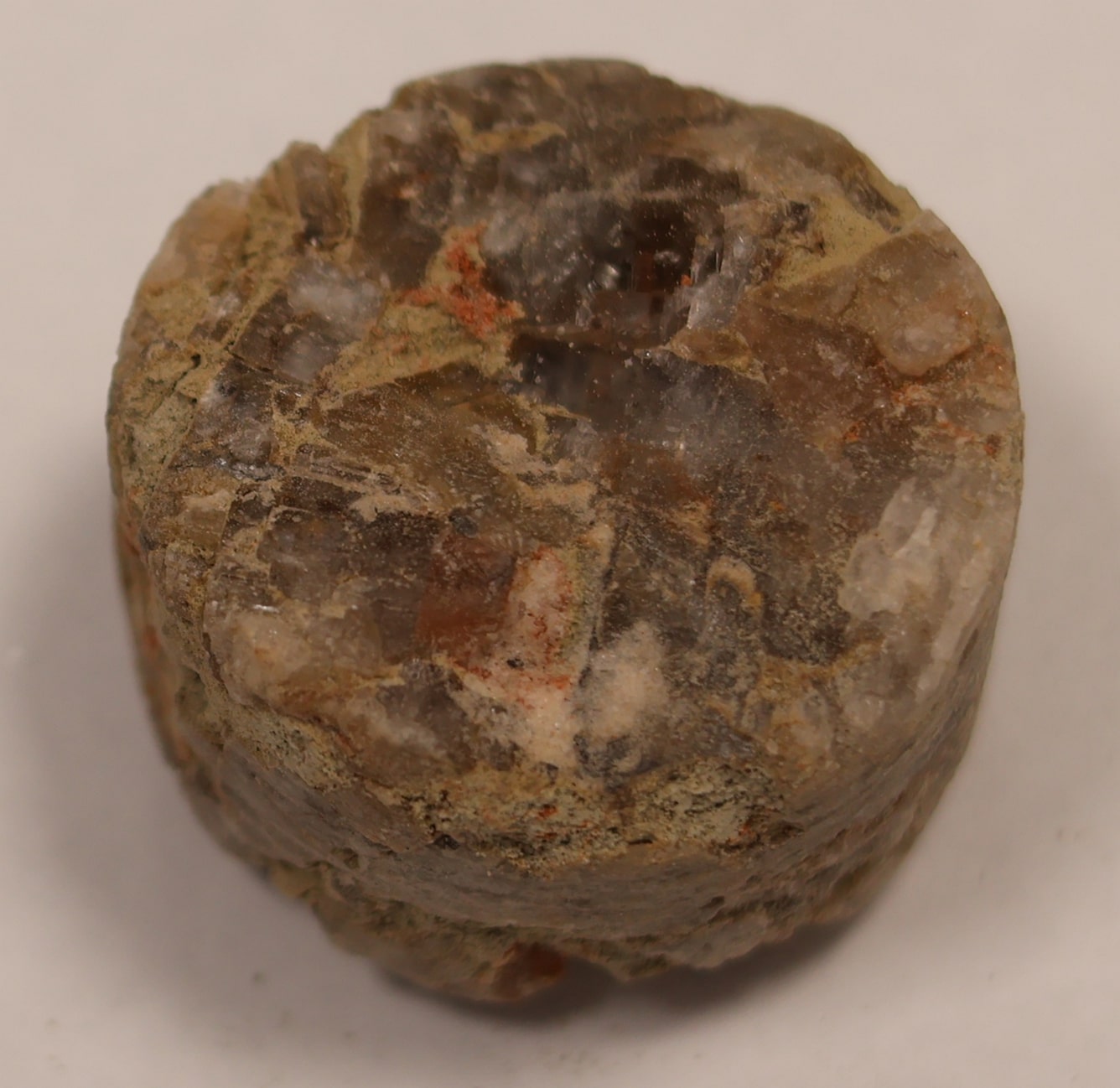}
\caption{Halite 4} \label{fig:Rocks_g}
\end{subfigure}
\begin{subfigure}{0.32\textwidth}
\includegraphics[width=\linewidth,angle=0]{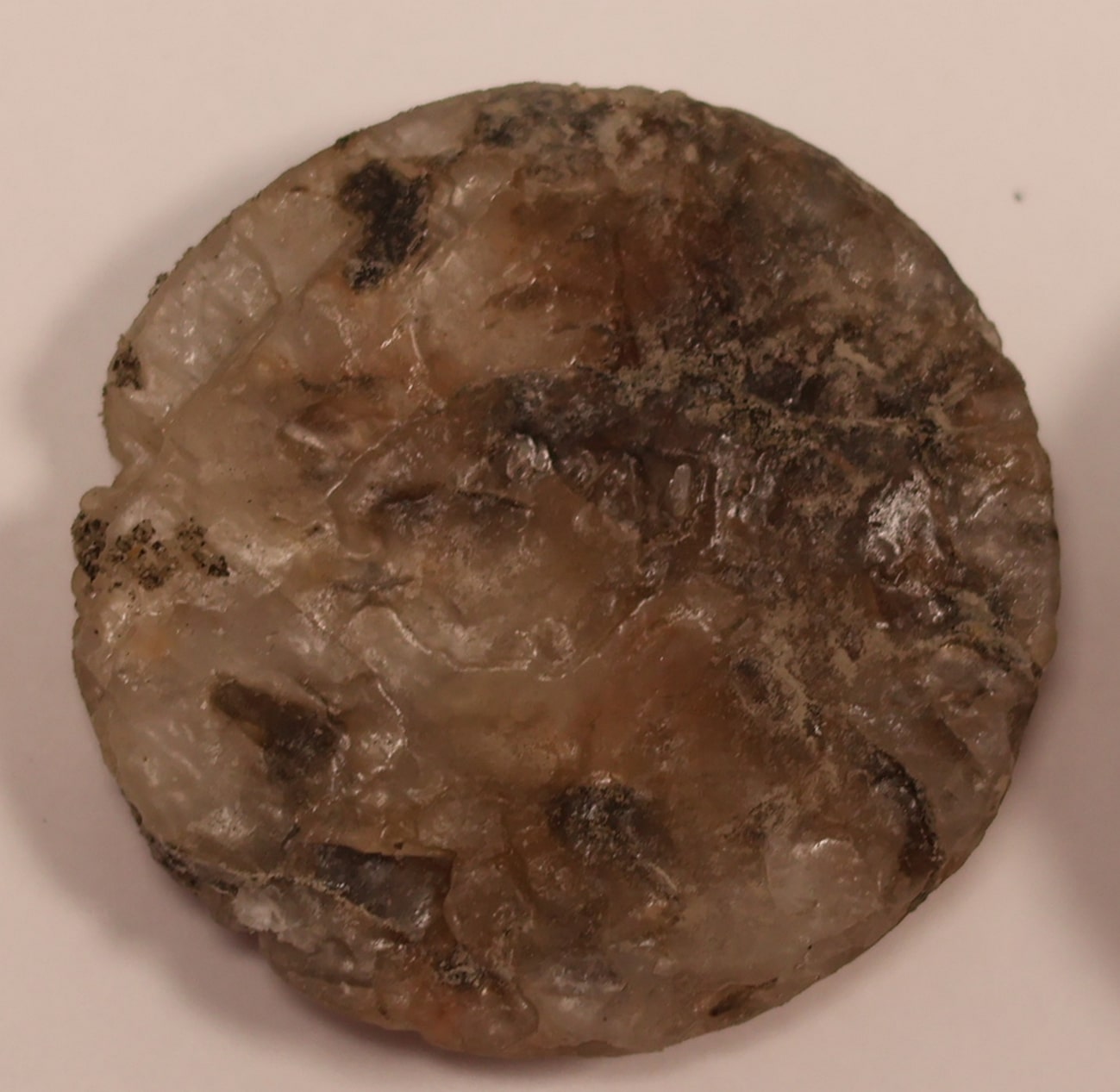}
\caption{Halite 9} \label{fig:Rocks_i}
\end{subfigure}
    \caption{Photographs of examples of the different types of rock samples measured, as listed in \autoref{tab:samples}. The polyhalite samples are approximately 150~mm in length and 50~mm in diameter. All other rock types have approximate sample dimensions of 100~mm in length and 50~mm in diameter. Details of their compositions are given in Section~\ref{sec:boulby_results}.}
    \label{fig:rocks}
\end{figure}

The samples vary widely in characteristics such as composition, crystal size and density. To aid the sample descriptions, \autoref{fig:rocks} displays images of the measured rock types. The primary components of potash (sylvinite), in roughly equal amounts, are sylvite (KCl), which is the mineral form of potassium chloride, and halite (NaCl). There tend to be many impurities disseminated throughout potash, rendering it relatively unclean, particularly with low-grade (LG) potash which is categorised separately here. Also high in potassium is CPH as it is mainly clean halite with varying amounts of sylvite and some silt impurities. FWHL is similar in composition to CPH, but the halite is less clean and has more silt deposits. Halites 3 and 4 are similar in composition, the bulk being halite with small fractions of potash and carnallite (MgCl$_2$·6(H$_2$O)), differing in granularity and crystal size. Salt polygons are made from relatively clean halite, with little sylvite, silt or anhydrite (CaSO$_4$) and are structurally comprised of large crystals. Consequently, the polygons have low levels of $^{40}$K and other NORM isotopes relative to the bands above. Halite 9 is a mixture of halitic anhydrite, anhydritic halite and silt, it is much lower in potassium than other samples with a higher halite content. Shale may also be present in some bands. The anhydrite band lacks halite and is low in potassium relative to other rock types, being further from the sylvite deposits. The traces of $^{238}$U and $^{235}$U are relatively high in anhydrite.

The polyhalite samples are from boreholes separate from those spanning all of the other bands of rock: one at a depth of $\sim\,$1100~m and two near a depth of $\sim\,$1300~m, the latter being closer in depth and location to where a future laboratory may be built. Polyhalite (K$_2$Ca$_2$Mg(SO$_4$)$_4$·2H$_2$O) is a very uniform material with a small grain size and high purity. The $^{40}$K content is very high, similar to that of potash. However, other contaminants are scarce, resulting in very low levels of $^{232}$Th and relatively low levels of $^{238}$U. A summary of information on the different rock types can be found in~\ref{sec:apx:boulby_rocks}.

\subsection{Discussion of sample measurements}
\label{sec:boulby_disc}
Previous measurements of the background radioactivity at the 1100~m level in Boulby were carried out in 2006~\cite{2013_BUL}. Halite and mudstone samples were taken from the walls just outside the laboratory and measured with an HPGe detector. In situ measurements, next to the same walls where samples were collected from, were also performed with an HPGe detector. Although the location of the laboratory is about 1~km away from the borehole sources of our samples, and it was either not known or not disclosed exactly which type of halite they measured, the magnitude of the radioactivity can be compared. From halite samples (with some sylvite impurities), the average decay rate of $^{232}$Th was measured as 0.6~Bq~kg$^{-1}$ and the average decay rate of $^{238}$U was measured as 0.4~Bq~kg$^{-1}$. From the in situ measurements (of a mixture of halite and mudstone in the walls) the average decay rate of $^{232}$Th was measured as 1.88~Bq~kg$^{-1}$ and the average decay rate of $^{238}$U was measured as 1.63~Bq~kg$^{-1}$. Similarly low activity concentrations characterise our halite samples, but still with a large spread, of $^{232}$Th: $0.161~-~4.262~$~Bq~kg$^{-1}$, and $^{238}$U: $0.145~-~2.323$~Bq~kg$^{-1}$. Note that these ranges do not include measurements from potash, anhydrite or polyhalite as the majority of their compositions are not halite.

The rock composition in the simulation mirrors that of Boulby mine at the 1100~m level, which is predominantly halite. To ensure the simulation yields consistent results with different initial rock compositions, we reran the simulation with polyhalite instead of pure NaCl as they have vastly different elemental compositions. The results of this test, displayed in~\ref{sec:apx:salt_v_poly}, show negligible differences between the shape of the two spectra. There is a spectrum-wide fractional difference of $\sim$23\,\% in raw counts. However, when the results are normalised to 1~Bq~kg$^{-1}$ it is clear that the difference in rock density is the determining factor in the raw counts difference (NaCl: 2.17~g~cm$^{-3}$, polyhalite: 2.78~g~cm$^{-3}$). Therefore, the simulation results can be used with any rock type regardless of composition.

We can now take the simulated results and normalise them to measurements of Boulby polyhalite since a potential future laboratory will be built in this rock-type band. The results are shown in \autoref{tab:samples_norm}. Since the rock strata undulate, there is a chance that the future laboratory may be partially built in rock that is not polyhalite. Therefore, the measurements of the many types of rock serve as a future reference for what thickness of shielding will be needed for a location-specific, rock gamma-ray background.
\vspace{-2mm}
\begin{center}
\normalsize
\def\arraystretch{1.23}%
\begin{longtable}[hbt!]{|c|c|c|c|c|}
\caption{Rates of events in the TPC with analysis cuts applied, normalised to measurements of polyhalite from Boulby mine (see \autoref{tab:samples}). Statistical uncertainties are quoted as described in the \autoref{tab:Rates} caption.}
\label{tab:samples_norm} \\
\hline
\multicolumn{1}{|c|}{} & \multicolumn{1}{c|}{\textbf{0 - 20~keV}} & \multicolumn{1}{c|}{\textbf{0 - 100~keV}} & \multicolumn{1}{c|}{\textbf{2408 - 2508 keV}} \\[-0.3cm]
\multicolumn{1}{|c|}{\textbf{Isotope}} & \multicolumn{1}{c|}{\textbf{Rate [year$^{-1}$]}} & \multicolumn{1}{c|}{\textbf{Rate [year$^{-1}$]}} & \multicolumn{1}{c|}{\textbf{Rate [year$^{-1}$]}} \\
\hline
\endfirsthead \hline 
\endlastfoot
$^{232}$Th & $\left(1.7_{-0.7}^{+1.3}\right)\times10^{-5}$ & $\left(1.62\pm0.26\right)\times10^{-4}$ & $\left(1.11 \pm 0.07\right)\times10^{-3}$ \\
\hline
$^{238}$U  & $\left(7.8_{-5.0}^{+9.0}\right)\times10^{-5}$ & $\left(5.7_{-1.4}^{+1.8}\right)\times10^{-4}$ & $\left(3.25\pm 0.13\right)\times10^{-2}$ \\
\hline
$^{40}$K & $\left(0_{-0}^{+9.5}\right)\times10^{-2}$ & $\left(0_{-0}^{+9.5}\right)\times10^{-2}$ & n/a \\
\end{longtable}
\end{center}

\section{Discussion}
\label{sec:discussion}
For dark matter searches, a background of $<\,$1~event per year is required to ensure a potential signal can be found. Further NR/ER discrimination can reduce the background that mimics NR events by more than two orders of magnitude. At such a low rate, the experiment could run for 10 years, with the background after discrimination remaining negligible ($<0.1$~events over 10 years), assuming the 99.5\,\% discrimination already achieved~\cite{LZfirstResults}.
The sensitivity requirement for 0νββ decay is even stricter, as signal events may be indistinguishable from background events (both being produced by electrons). Therefore, the background must be limited to $<1$ event over 10 years of data collection. Our findings (\autoref{tab:Rates}, columns 2-5) show that 4~m of water and scintillator on the top and sides, combined with 2~m on the bottom and a steel plate, is more than sufficient to reduce the gamma-ray background from rock for WIMPs to a negligible level, for an initial radioactivity in rock of about 1~Bq~kg$^{-1}$. For 0νββ decay search, the rate of background events for 1~Bq~kg$^{-1}$ of $^{232}$Th and $^{238}$U concentrations (\autoref{tab:Rates}, rows 6-7) suggests that the fiducial volume needs to be reduced compared to that typically used for WIMP searches. Depending on the concentration of radionuclides in the rock at the future site, the fiducial volume in the TPC needs to be adjusted to meet the background requirements for the 0νββ decay search.

The polyhalite samples from Boulby provide some insight into the radioactive background where a future laboratory may be constructed. Ideally, additional samples would be collected and measured to give a more accurate understanding of the surroundings. However, polyhalite is known to be fairly uniform in composition, so the results in \autoref{tab:samples_norm} can be used to give a general overview of the background. When the simulation results are normalised to the screening results of the polyhalite samples, the background event rates of $^{232}$Th and $^{238}$U are much lower than necessary for the sensitivity requirements in either of the 0νββ decay or WIMP ROIs. Therefore, the shielding thickness could be significantly reduced. Despite the rates from $^{40}$K decay being much higher, the $^{40}$K gamma-rays will not affect the 0νββ decay search, and the rate is still $<1$~event per year in the WIMP ROI.

Due to other sources of background, such as those from the detector materials themselves, the fiducial volume for 0νββ decay will be much smaller than the one defined above in Section 2. Given the much smaller fiducial volume for 0νββ decay due to the background from detector components and the sufficiently low background in the WIMP ROI, the water tank could be reduced in size which would consequently reduce the cost of construction. The simulated water shielding thickness is 3.5~m on the top and sides of the detector and 1.5~m underneath. The detector is also shielded by 0.5~m of GdLS on all sides and a 30~m thick steel plate beneath the water tank. Surfaces 1 - 5, which are detailed in \autoref{tab:surfaces}, are concentrically layered throughout the water shielding from just outside of the water tank to just outside the GdLS layer, as illustrated in \autoref{fig:geom_b}. 
To estimate the resultant factor of increase in the rate at different shielding thicknesses, we used the innermost attenuation coefficient (the attenuation from surface 4 to surface 5). For $^{232}$Th, we use the attenuation coefficient of the 2.615~MeV gamma-ray because after 3~m of shielding the spectrum is fully determined by the initial $^{208}$Tl line. For $^{238}$U, we use the attenuation coefficient of the 2.448~MeV gamma-ray because it is the highest energy gamma-ray in the decay chain and the largest contributor to the spectrum in the TPC after 3~m of shielding. Additionally, it is the only gamma-ray whose energy will affect the background in the 0νββ decay ROI. The flux ratio is calculated using the following equation:
\begin{align}\label{eq:flux_ratio}
    \frac{I}{I_{0}} = \exp{^{(-A x)}}
\end{align}
where $I$ is the flux at surface 5 and $I_0$ is the flux at surface 4 with $x=1$~m separating them. We find that A is 0.0439~cm$^{2}$~g$^{-1}$ for the flux ratio of 2.615~MeV gamma-rays and 0.0472~cm$^{2}$~g$^{-1}$ for the flux ratio of 2.448~MeV gamma-rays. Using these attenuation coefficients, we can apply \autoref{eq:flux_ratio} to calculate the increase in background rate when reducing water thickness. The reduction of water thickness is listed in the first column of \autoref{tab:increased_rates}. The relative increases in background rates are given in columns 2 and 3 for $^{232}$Th and $^{238}$U decay chains, respectively. Multiplying the original background event rates in the WIMP ROI (\autoref{tab:Rates}, column 3) by these rate increase factors, we gain new rates given in columns 4 and 5 of \autoref{tab:increased_rates}. These rates indicate that even with one metre less water shielding from all sides, the background provided by cavern gamma-rays still falls below sensitivity requirements for WIMP search. The thickness of water shielding and GdLS around the detector would be 3~m on the top and sides of the detector and 1~m beneath the detector, with the 30~cm thick steel plate at the base of the water tank. The volume of the water tank would decrease by 43.4\,\%, from 1216.7~m$^3$ to 688.2~m$^3$.

\begin{center}
\begin{minipage}{\textwidth}
\footnotesize
\def\arraystretch{1.2}%
\begin{longtable}[hbt!]{|c|c|c|c|c|}
\caption{Estimated event rates from 1~Bq kg$^{-1}$ of $^{232}$Th and $^{238}$U in the TPC, depositing energy in the WIMP region of interest (0~-~20~keV), are shown for different water shielding thicknesses. Column 1 specifies how much the shielding is reduced from the top, base, and sides of the simulated water tank. The rate increases show how much the event rates (see \autoref{tab:Rates}) rise as shielding thickness decreases from the original simulated thickness, with the updated rates presented in columns 4 and 5. Statistical uncertainties are quoted as described in the \autoref{tab:Rates} caption.}
\label{tab:increased_rates}
\\
\hline
\multicolumn{1}{|c}{\textbf{Thickness}} & \multicolumn{1}{|c|}{\textbf{Rate increase}} & \multicolumn{1}{|c|}{\textbf{Rate increase}} & \multicolumn{1}{c|}{\textbf{$^{232}$Th Rate}} & \multicolumn{1}{c|}{\textbf{$^{238}$U Rate}}\\[-0.12cm]
\multicolumn{1}{|c}{\textbf{reduction [cm]}} & \multicolumn{1}{|c|}{\textbf{$^{232}$Th}} & \multicolumn{1}{|c|}{\textbf{$^{238}$U}} & \multicolumn{1}{c|}{\textbf{[year$^{-1}$(Bq/kg)$^{-1}$]}} & \multicolumn{1}{c|}{\textbf{[year$^{-1}$(Bq/kg)$^{-1}$]}}\\
\endfirsthead
\endlastfoot
\hline
0 & 1 & 1 & $\left(1.5_{-0.6}^{+1.1}\right)\times10^{-3}$ & $\left(2.2_{-1.4}^{+2.5}\right)\times10^{-4}$ \\
\hline
25  & 3 & 3.3 & $\left(4.6_{-1.9}^{+3.2}\right)\times10^{-3}$ & $\left(7.3_{-4.6}^{+8.2}\right)\times10^{-4}$ \\
\hline
50 & 9.0 & 10.6 & $\left(1.4_{-0.6}^{+1.0}\right)\times10^{-2}$ & $\left(2.4_{-1.5}^{+2.7}\right)\times10^{-3}$\\
\hline
75 & 27.0 & 34.4 & $\left(4.1_{-1.7}^{+2.9}\right)\times10^{-2}$ & $\left(7.7_{-4.9}^{+8.7}\right)\times10^{-3}$\\
\hline
100 & 80.9 & 111.9 & $0.12_{-0.05}^{+0.09}$ & $\left(2.5_{-1.6}^{+2.8}\right)\times10^{-2}$\\
\hline
\end{longtable}
\end{minipage}
\end{center}

\par Using these rate increases, the size of the fiducial volume necessary for sensitivity to 0νββ decay can be estimated for the corresponding decreases in shielding. The results of this analysis are in \autoref{tab:FV_rates} and displayed in \autoref{fig:FV_min}.

\begin{center}
\footnotesize
\def\arraystretch{1.4}%
\begin{longtable}[ht!]{|c|c|c|c|c|}
\caption{Rates of events in the TPC that deposit energy in the 0νββ decay region of interest ($\pm\,50$~keV around the Q-value) from 1~Bq kg$^{-1}$ of $^{232}$Th and $^{238}$U, multiplied by their respective rate increase factors listed in \autoref{tab:increased_rates}. The fiducial volume was gradually decreased until the rate fell to less than 1 event per 10 years for each reduction in shielding thickness, from each of the top, sides, and base of the water tank. Statistical uncertainties are quoted as described in the \autoref{tab:Rates} caption.}
\label{tab:FV_rates}
\\
\hline
\multicolumn{1}{|c}{\textbf{Thickness}} & \multicolumn{1}{|c}{\textbf{$^{232}$Th rate}} & \multicolumn{1}{|c|}{\textbf{$^{232}$Th FV}} & \multicolumn{1}{|c|}{\textbf{$^{238}$U rate}} & \multicolumn{1}{c|}{\textbf{$^{238}$U FV}} \\[-0.2cm] 
\multicolumn{1}{|c}{\textbf{reduction [cm]}} & \multicolumn{1}{|c}{\textbf{[year$^{-1}$~(Bq/kg)$^{-1}$]}} & \multicolumn{1}{|c|}{\textbf{[tonne]}} & \multicolumn{1}{|c|}{\textbf{[year$^{-1}$~(Bq/kg)$^{-1}$]}} & \multicolumn{1}{c|}{\textbf{[tonne]}} \\
\endfirsthead
\endlastfoot
\hline
0 & $(9.69\pm0.61)\times10^{-2}$ & 39.3 & $(9.19\pm0.32)\times10^{-2}$ &  59.6 \\
\hline
25 & $\left(9.88\pm1.07\right)\times10^{-2}$  & 32.9 & $(9.90\pm0.60)\times10^{-2}$ & 51.2 \\
\hline
50 & $\left(9.99\pm1.86\right)\times10^{-2}$  & 26.5 & $(9.24\pm1.05)\times10^{-2}$ & 45.1 \\
\hline
75 & $\left(9.30_{-2.80}^{+3.92}\right)\times10^{-2}$  & 22.9 & $(9.64\pm1.93)\times10^{-2}$ & 37.8 \\
\hline
100 & $\left(9.30_{-5.89}^{+7.13}\right)\times10^{-2}$  & 19.1 & $\left(8.78_{-3.45}^{+4.14}\right)\times10^{-2}$ & 30.9 \\
\hline
\end{longtable}
\end{center}

\begin{figure}[htbp]
\begin{subfigure}{.49\textwidth}
    \centering
    \centering\includegraphics[width=1\textwidth,angle=0]{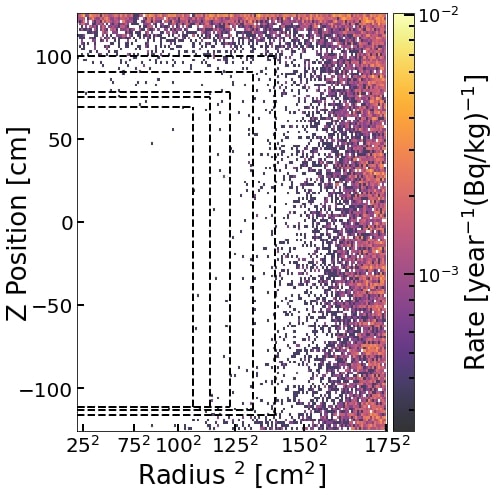}
        \caption{}
    \label{fig:FVs_Th}
    \end{subfigure}
    \begin{subfigure}{.49\textwidth}
    \centering
    \centering\includegraphics[width=1\textwidth,angle=0]{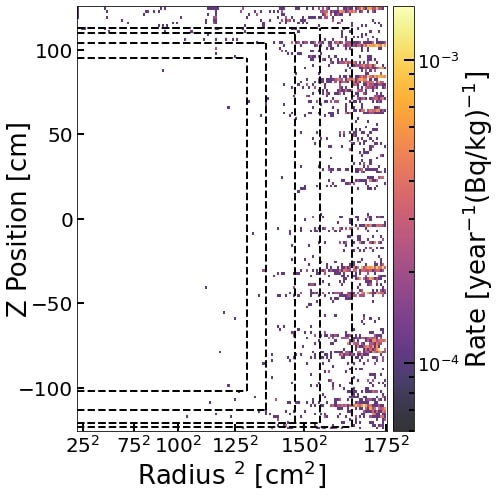}
    \caption{}
    \label{fig:FVs_U}
    \end{subfigure}
    \caption{Colour maps of positions of background events from (a) $^{232}$Th and (b) $^{238}$U (using the same data plotted as that in~\cref{fig:MS_cut_Qd,fig:MS_cut_Qf}, respectively). The fiducial volume was decreased gradually to maintain a rate of less than 1 event per 10 years in the TPC as the water shielding was reduced from the top, sides and base by values in \autoref{tab:FV_rates}. This means up to a 51.3\% decrease in LXe mass for $^{232}$Th and up to a 48.2\,\% decrease in LXe mass for $^{238}$U.}
\label{fig:FV_min}
\end{figure}

\section{Conclusions}
\label{sec:conc}
To shield the next-generation dark matter detector from the surrounding rock's gamma-ray and neutron background, 4 metres of water and GdLS shielding on the top and sides, and 2 metres on the bottom with a steel plate base, has been shown to be sufficient for WIMP search and 0νββ decay detection, based on a full GEANT4 Monte Carlo simulation. 
In the WIMP search ROI, 0~-~20~keV, $^{40}$K, $^{238}$U and $^{232}$Th each contribute $<<1$ event per year to the background assuming their activity of 1~Bq~kg$^{-1}$. Similarly, there was $<1$ event per 10 years in the 0νββ decay ROI, 2.408~-~2.508~MeV, from $^{238}$U and $^{232}$Th, each having an activity of 1~Bq~kg$^{-1}$.

Using attenuation factors of the highest energy gamma-rays in the $^{232}$Th and $^{238}$U decay chains, we estimate that with 1~m less water shielding on all sides of the detector, the TPC event rates in the WIMP ROI increase from $\left(1.5_{-0.6}^{+1.1}\right)\times10^{-3}$ to $0.12_{-0.05}^{+0.09}$~year$^{-1}$~(Bq/kg)$^{-1}$ and $\left(2.2_{-1.4}^{+2.5}\right)\times10^{-4}$ to $\left(2.5_{-1.6}^{+2.8}\right)\times10^{-2}$~year$^{-1}$~(Bq/kg)$^{-1}$, for $^{232}$Th and $^{238}$U, respectively. This decrease in shielding necessitates stricter fiducial volumes for the 0νββ decay ROI to remain below 1~event per 10 years. Specifically, fiducial xenon masses for $^{232}$Th and $^{238}$U background rates would decrease from 39.3~to~19.1~tonnes and 59.6~to~30.9~tonnes, respectively. Other sources of background such as those originating from the detector materials will reduce the fiducial volume further.

With the potential of housing the future detector in Boulby Mine, different rock types from various geology bands were measured to construct a guide to the radioactive backgrounds that could be encountered. The polyhalite, although high in $^{40}$K, is low in both $^{232}$Th and $^{238}$U with specific radioactivities of, respectively, $\left(1.14\pm0.37\text{ \small(stat.)}\pm0.06\text{ \small(sys.)}\right)\times10^{-2}$~Bq~kg$^{-1}$ and $0.354\pm0.008\text{ \small(stat.)}\pm0.042\text{ \small(sys.)}$~Bq~kg$^{-1}$, which makes the location a suitable site for a high-sensitivity dark matter experiment.

\section*{Acknowledgements}
We would like to thank STFC for funding this project and the ICL Mining Company for access to the Boulby mine, rock samples, and two of their geologists, P. Edey and D. Webb, who provided us with materials to understand the geology of Boulby. We would also like to thank the whole team from the Boulby Underground Laboratory.

\bibliographystyle{elsarticle-num} 

\newpage
\appendix
\section*{Appendices}
\section{}
\label{sec:apx:std_dev}
\begin{figure}[ht!] 
\centering
\begin{subfigure}{0.3\textwidth}
\includegraphics[width=\linewidth,angle=0]{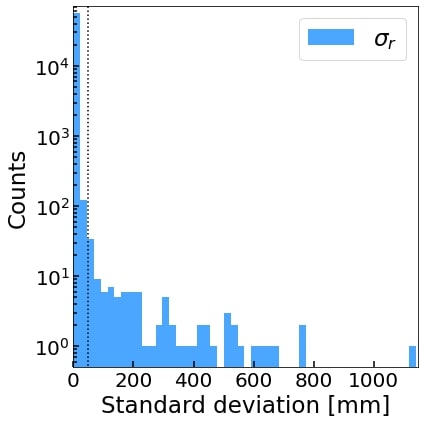}
\setcounter{figure}{1}
    \renewcommand\thefigure{\Alph{figure}\arabic{figure}}
\vspace{-8mm}
\caption{} \label{fig:Th100r}
\end{subfigure}
\begin{subfigure}{0.3\textwidth}
\includegraphics[width=\linewidth,angle=0]{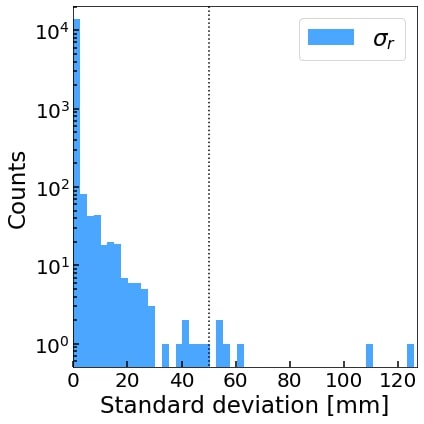}
\setcounter{subfigure}{2}
\vspace{-8mm}
\caption{} \label{fig:U100r}
\end{subfigure}
\begin{subfigure}{0.3\textwidth}
\includegraphics[width=\linewidth,angle=0]{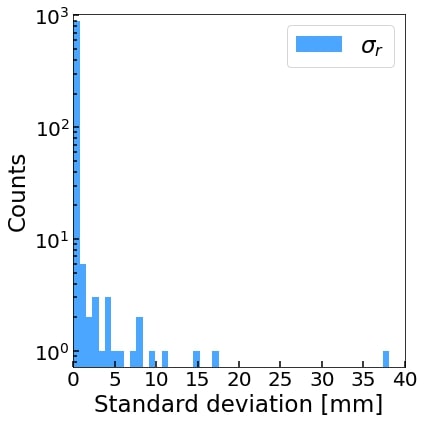}
\setcounter{subfigure}{4}
\vspace{-8mm}
\caption{} \label{fig:K100r}
\end{subfigure}
\vspace{-3mm}
\medskip
\begin{subfigure}{0.3\textwidth}
\includegraphics[width=\linewidth,angle=0]{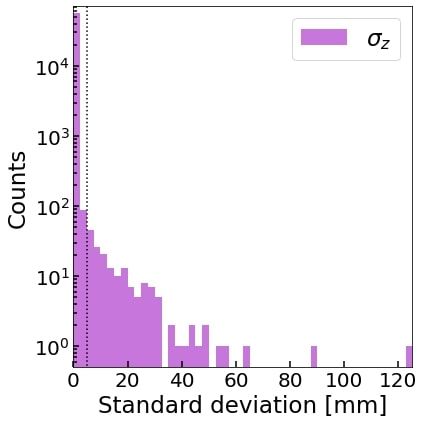}
\setcounter{subfigure}{1}
\vspace{-8mm}
\caption{} \label{fig:Th100z}
\end{subfigure}
\begin{subfigure}{0.3\textwidth}
\includegraphics[width=\linewidth,angle=0]{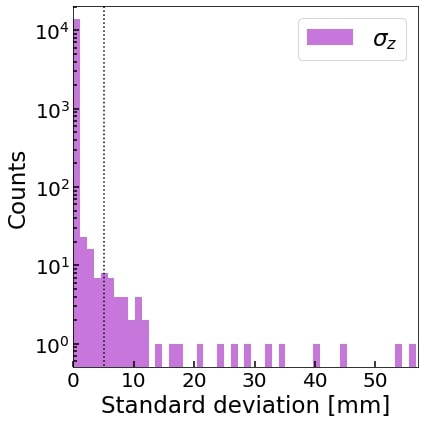}
\setcounter{subfigure}{3}
\vspace{-8mm}
\caption{} \label{fig:U100z}
\end{subfigure}
\begin{subfigure}{0.3\textwidth}
\includegraphics[width=\linewidth,angle=0]{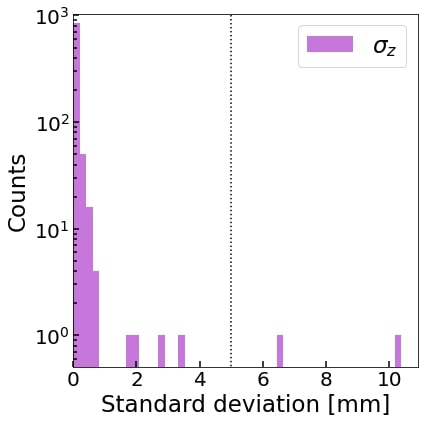}
\setcounter{subfigure}{5}
\vspace{-8mm}
\caption{} \label{fig:K100z}
\end{subfigure}
    \caption{Radial (top) and $z$-coordinate (bottom) standard deviations of $^{232}$Th (a and b), $^{238}$U (c and d), and $^{40}$K (e and f) events in the TPC in the energy range 0~-~100~keV.}
    \label{fig:std_dev}
\end{figure}

\begin{figure}[ht!] 
\centering
    \begin{subfigure}{.35\textwidth}
    \centering
    \centering\includegraphics[width=1\textwidth,angle=0]{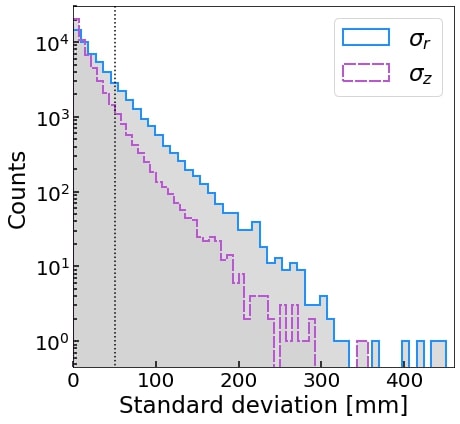}
        \caption{}
    \label{fig:ThQ}
    \end{subfigure}
    \begin{subfigure}{.35\textwidth}
    \centering
    \centering\includegraphics[width=1\textwidth,angle=0]{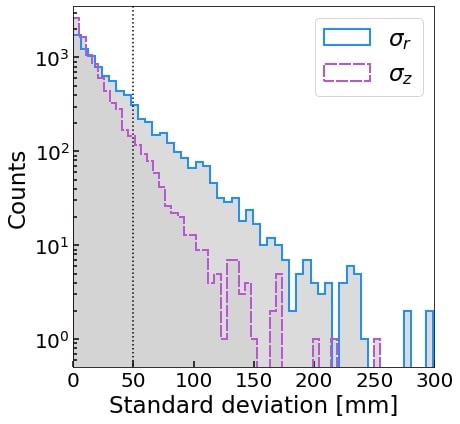}
        \caption{}
    \label{fig:UQ}
    \end{subfigure}
    \caption{Radial and $z$-coordinate standard deviations of (a) $^{232}$Th and (b) $^{238}$U events in the TPC in the energy range 2408~-~2508~keV, $\pm~50$~keV around the $0\nu\beta\beta$ decay Q-value.}
    \label{fig:std_dev_Q}
\end{figure}

\newpage
\section{}
\label{sec:apx:sample}

\begin{figure}[ht!]
    \centering 
    \setcounter{figure}{0}
    \renewcommand\thefigure{B\arabic{figure}}
    \includegraphics[width=0.8\linewidth]{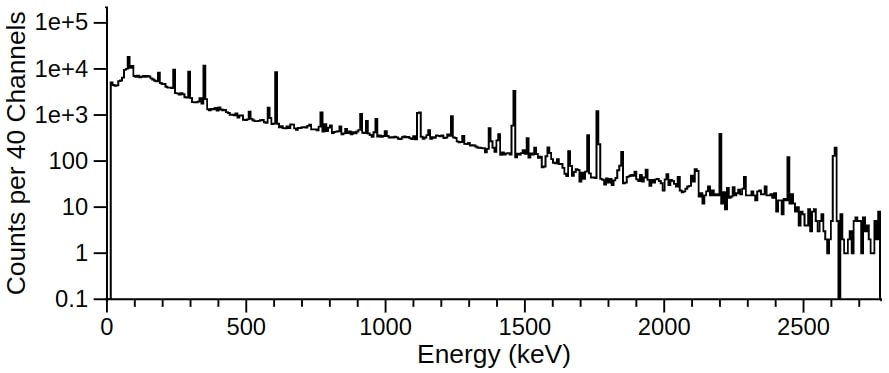}
    \caption{Energy spectrum of an anhydrite sample as measured by the BEGe detector, ``Chaloner'', in the Boulby UnderGround Screening facility. The sample had a mass of 0.828~kg and was measured for 159~hours.}
    \label{fig:my_label_b1}
\end{figure}

\section{} \label{sec:apx:boulby_sys_uncert}
\begin{itemize}
    \item Uncertainty on weight: $\pm~1$~g
    \item Uncertainty on branching ratio is peak dependent and taken from Ref.~\cite{Gilmore}.
    \item Uncertainty on efficiency with respect to a sample's density, length and diameter values entered into LabSOCS is peak dependent:
    \begin{itemize}
        \item diameter: $\pm~1$~mm
        \item length: $\pm~5$~mm
        \item density: $\pm~10\%$
    \end{itemize}
    \item Uncertainty on LabSOCS software efficiency calculations at $1\sigma$~\cite{labsocs_spec}:
    \begin{itemize}
        \item Peak energies 0 - 50 keV: $<20$\%
        \item Peak energies 50 - 400 keV: 7 - 11\% 
        \item Peak energies $>400$ keV: 4 - 5\%
    \end{itemize}
\end{itemize}
To be conservative, the upper end of the uncertainty ranges were used. Each peak in every sample has an efficiency calculated by LabSOCS. The uncertainty on this efficiency was calculated according to the peak's energy. Total systematic uncertainty has been obtained by summing individual uncertainties in quadrature.

\newpage
\section{}
\label{sec:apx:boulby_rocks}
\begin{table}[ht!]
\setcounter{table}{0}
    \renewcommand\thetable{D\arabic{table}}
    \caption{Information about the mineral composition of the various rock types that were measured with high-purity germanium detectors and discussed in Section~\ref{sec:boulby_results}.}\label{tab:rock_info}
    \centering
    \begin{tabular}{|c|c|}
    \hline
       \textbf{Rock type}  & \textbf{Characteristics} \\
       \hline
       Polyhalite  &  K$_2$Ca$_2$Mg(SO$_4$)$_4$·2H$_2$O\\
       \hline
       Salt polygons  &  Mainly NaCl, minor amounts of silt and KCl impurities. \\
       \hline
       Potash  &  Roughly equal parts KCl and NaCl, many impurities.\\
       \hline
       Low-grade potash & Similar to higher-grade potash, but with more impurities.\\
       \hline
       Footwall halite (FWHL)  &  Mainly NaCl with some KCl and silt deposits. \\
       \hline
       Clear pink halite (CPH)  & Mainly NaCl with some KCl, fewer silt deposits than FWHL.\\
       \hline
       Anhydrite  &  CaSO$_4$ \\
       \hline
       Halite 3  &  Mainly NaCl, some KCl and MgCl$_2$·$6($H$_2$O). Small crystal size.\\
       \hline
       Halite 4  &  Mainly NaCl, some KCl and MgCl$_2$·$6($H$_2$O)). Large crystal size. \\
       \hline
       Halite 9  &  Mixture of NaCL and CaSO$_4$.\\
       \hline
    \end{tabular}
\end{table}

\section{}
\label{sec:apx:salt_v_poly}
\begin{figure}[ht!]
    \centering 
    \setcounter{figure}{0} 
    \renewcommand\thefigure{E\arabic{figure}}
    \begin{subfigure}{.43\textwidth}
    \centering
    \includegraphics[width=1
\textwidth,angle=0]{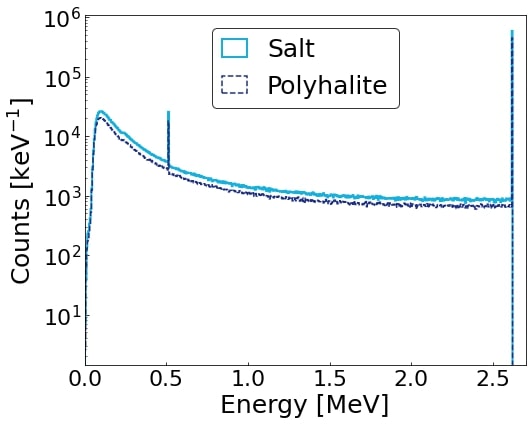}
    \caption{}
    \label{fig:my_label_e1a}
    \end{subfigure}
    \begin{subfigure}{.45\textwidth}
    \centering
    \includegraphics[width=1
\textwidth,angle=0]{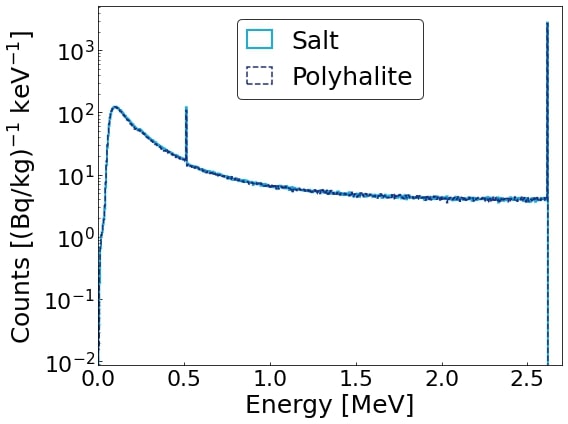}
    \caption{}
    \label{fig:my_label_e1b}
    \end{subfigure}
    \caption{Energy spectra of gamma-rays at the 1st tracking surface, 10~cm outside of the water tank. $3.6\times10^8$ gamma-rays, each with an energy of 2.615~MeV,  were generated from salt and polyhalite rock separately to check for differences between the resultant spectra. The number of gamma-rays that reached the 1st surface from salt and polyhalite is 10702179 and 8170331, respectively. (a) displays the difference in the raw counts showing a clear distinction between the datasets, and (b) demonstrates the effect of normalising the counts to 1~Bq~kg$^{-1}$, indicating that the difference between spectra in \autoref{fig:my_label_e1a} is determined by the rock density.}
\end{figure}

\end{document}